\documentclass[pra,aps,twocolumn,superscriptaddress,showpacs]{revtex4}
\usepackage{graphicx}
\usepackage{amsfonts}
\usepackage{amssymb}
\usepackage{amsmath}
\usepackage{bm}
\usepackage{bbm}
\makeindex
\DeclareMathAlphabet\mathbfcal{OMS}{cmsy}{b}{n}

\renewcommand{\r}{\textbf{r}}
\newcommand{\Y}{\textbf{Y}}
\newcommand{\E}{\textbf{E}}
\newcommand{\J}{\textbf{J}}
\newcommand{\Q}{\textbf{Q}}
\renewcommand{\H}{\textbf{H}}

\renewcommand{\S}{\textbf{S}}
\renewcommand{\Re}{\mathrm{Re}}
\renewcommand{\Im}{\mathrm{Im}}

\newcommand{\er}{\textbf{e}_r}
\newcommand{\ephi}{\textbf{e}_\varphi}
\newcommand{\etheta}{\textbf{e}_\theta}
\newcommand{\iE}{\mathbfcal{E}}
\newcommand{\G}{\hat{\textbf{G}}}
\newcommand{\g}{\hat{\mathbfcal{G}}}
\newcommand{\e}{\mathbfcal{E}}

\newcommand{\one}{\hat{\bm 1}}

\newcommand{\grad}{\pmb{\nabla}}
\newcommand{\eps}{\hat{{\pmb{\varepsilon}}}}
\newcommand{\rot}{\pmb{\nabla}\times}
\newcommand{\Eq}[1]{Eq.\,(\ref{#1})}
\newcommand{\Eqs}[2]{Eqs.\,(\ref{#1})-(\ref{#2})}

\newcommand{\Fig}[1]{Fig.\,\ref{#1}}
\newcommand{\Sec}[1]{Sec.\,\ref{#1}}
\newcommand{\App}[1]{Appendix\,\ref{#1}}
\newcommand{\ssc}{\sigma^{\rm sca}}
\newcommand{\sscl}{\sigma^{\rm sca}_l}

\newcommand{\LL}{\hat{\cal{L}}}
\newcommand{\GF}{\hat{\mbox{\boldmath ${\cal G}$}}}

\newcommand{\br}{\mathbf{r}}

\newcommand{\bs}{\boldsymbol{\sigma}}
\newcommand{\En}{\mathbf{E}_n}

\newcommand{\kmax}{k_{\rm max}}

\newcommand{\bc}{\mathbf{c}}
\newcommand{\mk}{\mathbbm{k}}
\newcommand{\mM}{\mathbbm{M}}
\newcommand{\mD}{\mathbbm{D}}

\newcommand{\be}{\begin{equation}}
\newcommand{\ee}{\end{equation}}
\newcommand{\bea}{\begin{eqnarray}}
\newcommand{\eea}{\end{eqnarray}}

\begin{document}
\title{Resonant-state expansion of three-dimensional open optical systems: Light scattering}
\author{S.\,V. Lobanov}\email{LobanovS@cardiff.ac.uk}
\altaffiliation[Present address: ]{School of Medicine, Cardiff University, Cardiff CF24 4HQ, United Kingdom}
\affiliation{School of Physics and Astronomy, Cardiff University, Cardiff CF24 3AA, United Kingdom}
\author{W. Langbein}
\affiliation{School of Physics and Astronomy, Cardiff University, Cardiff CF24 3AA, United Kingdom}
\author{E.\,A. Muljarov}
\affiliation{School of Physics and Astronomy, Cardiff University, Cardiff CF24 3AA, United Kingdom}
\begin{abstract}
A rigorous method of calculating the electromagnetic field, the scattering matrix, and scattering cross-sections of an arbitrary finite three-dimensional optical system described by its permittivity distribution is presented. The method is based on the expansion of the  Green's function into the resonant states of the system. These can be calculated by any means, including the popular finite element and finite-difference time-domain methods. However, using the resonant-state expansion with a spherically-symmetric analytical basis, such as that of a homogeneous sphere, allows to determine a complete set of the resonant states of the system within a given frequency range. Furthermore, it enables to take full advantage of the expansion of the field outside the system into vector spherical harmonics, resulting in simple analytic expressions. We verify and illustrate the developed approach on an example of a dielectric sphere in vacuum, which has an exact analytic solution known as Mie scattering.
\end{abstract}
\date{\today}
\maketitle
\section{Introduction}

Resonances are at the heart of many physical  phenomena. In particular, they determine features in optical spectra such as the scattering and extinction cross-sections. Mathematically, resonances and their contribution to physical observables can be elegantly and rigorously described using the concept of resonant states (RSs)~\cite{SiegertPR39,Calderon1976b}.
The wave functions of RSs are generally not square integrable and therefore the usual definitions of normalization and orthogonality cannot be applied to them.
Several procedures for normalization of RSs in quantum mechanics were suggested, which include the Zel'dovich regularization~\cite{ZelDovich1961}, the use of the pole structure of the Green's function (GF) of the wave equation~\cite{More1971,More1973,Calderon1976,Calderon1976b}, as well as introducing a non-Hermitian scalar product~\cite{Shnol1971}.
Notably, the correct normalization of electromagnetic RSs was not known in the literature until recently, and normalizations used so far were approximate, as was clarified in~\cite{MuljarovPRB16a,MuljarovPRA17}. The correct normalization of RSs in electrodynamics, which takes into account the vectorial character of electromagnetic waves, was proposed in~\cite{MuljarovEPL10}, supplemented by a proof in \cite{DoostPRA14}, and recently generalized to systems with dispersion~\cite{MuljarovPRB16} and photonic crystal structures~\cite{WeissPRL16,WeissPRB17}. It has allowed to formulate an exact theory of the Purcell effect~\cite{MuljarovPRB16a}. This exact normalization of RSs is at the heart of the resonant-state expansion (RSE) method~\cite{MuljarovEPL10,DoostPRA14}, which is extended in the present work to the calculation of the scattering properties of open optical systems.

The concept of RSs is a powerful tool in treating open systems.
Similar to the standard Rayleigh-Schr\"odinger perturbation theory in Hermitian quantum mechanics~\cite{LandauLifshitz_3}, a non-Hermitian perturbation theory, describing changes in RSs due to small perturbations in the potential, can be constructed~\cite{More1971,More1973}.
Furthermore, it has been shown that the RSs of an open system form a complete set inside it~\cite{Calderon1976,Bang1978} and therefore can be used as a basis for expansions.
In particular, by projecting the wave equation onto the space of the RSs of a basis system, one can find RSs of a modified system.
This approach was  suggested in nuclear physics~\cite{Bang1978}, though never applied.
Recently, a similar idea has been implemented  in electrodynamics, leading to the formulation of the RSE~\cite{MuljarovEPL10,DoostPRA14,WeissPRL16,ArmitagePRA14,Lobanov2017}. The RSE is capable of treating accurately both small and large perturbations of an optical system.
Furthermore, the RSE has been recently extended~\cite{MuljarovPRB16} to systems with an arbitrary frequency dispersion of the permittivity described by a generalized Drude-Lorentz model~\cite{SehmiPRB17}.

A core quantity which contains the full information about an optical system is its dyadic GF.
This GF, being the solution of Maxwell's wave equation with a delta source term, describes the radiation of an oscillating point dipole, or a quantum emitter in the weak coupling limit~\cite{Lobanov11}. In general, the GF provides access to all physical observables, including the optical scattering matrix~\cite{Tikhodeev2002,Lobanov15b} ($S$-matrix) and the scattering cross-section, the electromagnetic near and far-field distributions, as well as the total radiation intensity and the Purcell factor~\cite{Purcell1946,Lobanov12,MuljarovPRB16a}.

While for specific systems the GF can be determined analytically, finding the GF in a general case is a challenging computational task.
It has been recently shown~\cite{MuljarovEPL10,DoostPRA14} that using the properly normalized RSs, the GF inside any optical system can be constructed efficiently using its spectral representation, also known as the Mittag-Leffler (ML) expansion. On the other hand, it has been also demonstrated in numerous publications (see e.g. Ref.~\cite{WeissJOSAA11}) that the $S$-matrix of an optical system has properties similar to those of the GF, namely, the poles of the $S$-matrix in the complex frequency plane are also determined by the RSs. While the information on the GF contained in the $S$-matrix is partial, being limited to external excitation, it completely describes the light scattered by the system. Therefore the problem of expressing the $S$-matrix in terms of the RSs reduces to establishing the link between the $S$-matrix and the GF, which is a key point of this paper.

To best of our knowledge, there has been only one report in the literature establishing a link between the $S$-matrix and the GF of a wave equation~\cite{ChaosCadorJPA10}; in that work a two-dimensional scalar Schr\"odinger wave equation is considered. Recently, a consistent phenomenological approach to presenting the $S$-matrix as a superposition of different isolated resonances has been suggested~\cite{AlpeggianiPRX17}. This approach is using the far-field information about the RSs of a system and demonstrates a limited accuracy. A similar method has been discussed earlier in~\cite{PerrinOE16}. Another approach, based on a Weierstrass factorization, has been recently suggested~\cite{GrigorievPRA13}. It is based on an elegant way of calculating the $S$-matrix in a spherically symmetric system using only the RS wave numbers, while the wave functions themselves are not required. It is not clear, however, if this approach can be generalized to scattering between different channels~\cite{Tikhodeev2002,Gippius2005,Maksimov14} which takes place in non-spherical systems.

In this paper, we present a rigorous scattering theory based on the concept of RSs, which determines the electromagnetic field, the $S$-matrix and any scattering properties of an arbitrary three-dimensional (3D) open optical system. The method is based on the following steps: (i) establishing the link between the $S$-matrix and the dyadic GF of an arbitrary 3D system and expressing them in the basis of vector spherical harmonics (VSHs); (ii) using the ML expansion of the GF to represent the $S$-matrix in terms of the RSs of the system; (iii) calculating the RSs via the RSE using a spherically symmetric basis system. We illustrate this theory on the analytically solvable example of a dielectric sphere in vacuum. We verify the approach by comparing the results with the exact analytic solution known as Mie theory~\cite{BohrenBook98}, demonstrating the convergence of the  $S$-matrix and the scattering cross-sections to the exact solution. We note that the approach to the light scattering developed in this work is expected to be suited also to open system in the vicinity of a planar interface, by following the treatment of flat boundaries suggested in~\cite{Fucile1997}.

We emphasize that the developed method can deal with general finite 3D systems. This includes systems of anisotropic shape \cite{DoostPRA14} and of a permittivity with dispersion~\cite{MuljarovPRB16}. Recently, the RSE formulation was extended to anisotropic, magnetic, chiral, and bi-anisotropic materials \cite{MuljarovOL18}, which includes the case of a biaxial sphere, solved previously using an extended Mie scattering formalism \cite{Liu2006}.
This might be useful for the understanding of non-reciprocity of magnetic surface plasmon~\cite{Liu2011,Yu2012} and the design of plasmonic~\cite{Chen2017a,Lobanov2017a} or chiral~\cite{Lobanov15} metasurfaces.

The paper is organized as follows. \Sec{Sec:Def_VSH} introduces VSHs and their properties required for the subsequent derivations. \Sec{Sec:GF_IP} makes the connection between incoming and outgoing fields and the GF of an open system. \Sec{Sec:IOEW} then uses this result to establish a link between the S-matrix and the GF. In \Sec{sec:RSE} we introduce the RSE and derive the normalization of the perturbed RSs used in the GF and the S-matrix. In \Sec{Sec:Results} we illustrate and verify the method on the analytically solvable example of a dielectric sphere in vacuum. Detailed derivations of solutions of the wave equation in the basis of the VSHs, as well as of the link between the GF and the S-matrix are presented in Appendices.

\section{Representation of fields in vector spherical harmonics }\label{Sec:Def_VSH}

In the present work, the RSs and the GF of a 3D open optical system are calculated using the RSE with a basis given by the analytically known RSs of a homogeneous sphere. Owing to the spherical symmetry, the resulting basis RSs are characterized by their azimuthal and total angular momentum quantum numbers. In the subsequent derivations we use VSHs to express the field outside of the system, because they form an orthogonal complete basis for expanding 3D vector fields on a unit sphere, and share the angular quantum numbers with the basis RSs, thus allowing for compact expressions. To prepare the use of VSHs, we provide here their definition and the mapping of vectors, tensors, and operators of the cartesian coordinate space $\r=(x,y,z)=(r\sin(\theta)\cos(\varphi), r\sin(\theta)\sin(\varphi),r\cos(\theta))$, with the spherical coordinates $(r, \theta, \varphi)$, onto the space of VSHs $\Y_{ilm}(\Omega)$, where $\Omega=(\theta,\varphi)$. The VSHs are defined as
\bea
\Y_{1lm}(\Omega) &=& \frac{1}{\sqrt{l(l+1)}} \r \times \grad Y_{lm}(\Omega)\,, \label{Eq:Y1}\\
\Y_{2lm}(\Omega) &=& \frac{1}{\sqrt{l(l+1)}} r \grad Y_{lm}(\Omega)\,, \label{Eq:Y2}\\
\Y_{3lm}(\Omega) &=& \frac{\r}{r} Y_{lm}(\Omega)\,,\label{Eq:Y3}
\eea
where $Y_{lm}(\Omega)$ are the {\em real } scalar spherical harmonics, defined in \Eq{Eq:Y} of \App{App:VSH} with $l$ and $m$ being the orbital and azimuthal quantum numbers, respectively. The VSHs are defined by \Eqs{Eq:Y1}{Eq:Y3} following Barrera {\it et. al.}~\cite{BarreraEJP85} as vector functions that are independent of the radius $r$, in contrast to their more traditional form given e.g. in \cite{StrattonBook41,BohrenBook98}. Different from Barrera {\it et. al.}, we use here real VSHs,
in order to satisfy the orthogonality condition of RSs without using the complex conjugate~\cite{DoostPRA14}, and choose their normalization to obey the orthonormality
\begin{equation}
\int \Y_{ilm}(\Omega)\cdot \Y_{i'l'm'}(\Omega)d \Omega=\delta_{ii'}\delta_{ll'}\delta_{mm'},
\end{equation}
where $d\Omega=\sin\theta d\theta d\varphi$ and $\delta_{nm}$ is the Kronecker delta.

Using the completeness of the VSHs, an arbitrary vector field, such as the electric field $\E(\r)$, can be expanded as
\begin{equation}
\E(\r) = \sum_{ilm} [{\bf E}_{lm}(r)]_i \Y_{ilm}(\Omega),\label{mapping}
\end{equation}
where the radially dependent expansion coefficient  ${\bf E}_{lm}(r)$ is a vector in the space of VSHs with given $l$ and $m$, and its $i$-th component $[{\bf E}_{lm}(r)]_i$ is the expansion coefficient corresponding to the $i$-th VSH $\Y_{ilm}(\Omega)$. In other words, \Eq{mapping} defines a mapping from real space to the VSH space, $\E(\r) \to \{{\bf E}_{lm}(r)\}$, where the curly brackets indicate the set of all $l,m$ components. This mapping is done in an equivalent way for tensors (for example, for the dielectric permittivity tensor $\eps(\r)\to \{\eps_{lm}^{l'm'}(r)\}$ or the dyadic GF $\GF(\r,\r';k)\to \{\GF_{lm}^{l'm'}(r,r';k)\}$, see~\App{App:VSH} for details) and for vector operators $\LL(\r;k)\to \{\LL_{lm}^{l'm'}(r;k)\}$.

Consider, for example, the Helmholtz operator, which generates the Maxwell wave equation:
\begin{equation}
\LL(\r;k)\equiv  k^2 \eps(\r) -\rot\rot \,,
\label{Helmholtz}
\end{equation}
where $k=\omega/c$ is the wave vector of the electromagnetic field in vacuum, which is given by the light frequency $\omega$ and speed $c$.
While the first term in \Eq{Helmholtz} is a tensor determined  by the structure of the optical system, the second term is block diagonal in the VSH basis,
\be
\left(\rot \rot \right)_{lm}^{l'm'}=\hat{L}_l\delta_{ll'}\delta_{mm'}\,,
\ee
where
\begin{equation}
\hat{L}_l=
\begin{pmatrix}
\displaystyle
\frac{l(l+1)}{r^2}-\frac{1}{r}\frac{d^2}{dr^2}r & 0 & 0\\
0 & \displaystyle-\frac{1}{r}\frac{d^2}{dr^2}r & \displaystyle\frac{\sqrt{l(l+1)}}{r}\frac{d}{dr} \\
0 & \displaystyle -\frac{\sqrt{l(l+1)}}{r^2}\frac{d}{dr}r & \displaystyle \frac{l(l+1)}{r^2}
\end{pmatrix},\label{Eq:L}
\end{equation}
as derived in \App{App:VSH}.
\section{Illuminated open optical system with known Green's function}\label{Sec:GF_IP}
Let us consider an optical system which is confined inside a sphere of radius $R$, surrounded by vacuum. The local time-independent permittivity tensor $ \eps(\r)$ outside the system is then $\eps(\r)=\one$ for $|\r|>R$, where $\one$ is the unit tensor. Note that systems embedded in a medium with an isotropic homogeneous real refractive index can be described in the same way, by simply rescaling $\eps(\r)$ and $k$.

Using the VSH representation, the Maxwell wave equation for the electric field,
\begin{equation}
\LL(\r;k)\E(\r)=0\,,
\label{ME}
\end{equation}
is mapped to the matrix differential equation
\begin{equation}
\sum_{l'm'}  \LL_{lm}^{l'm'}(r;k) {\bf E}_{l'm'}(r) =0\,,
\label{ME-mapped}
\end{equation}
see \App{App:VSH}. Here, $\LL_{lm}^{l'm'}(r;k)=k^2\eps_{lm}^{l'm'}(r)-\hat{L}_l\delta_{ll'}\delta_{mm'}$, and $\hat{L}_l$ is given by \Eq{Eq:L}. Similarly, the wave equation for the GF,
\begin{equation}
\LL(\r;k)\GF(\r,\r';k) = \one\delta(\r-\r')\,,
\end{equation}
is mapped to
\begin{equation}
\sum_{l''m''}  \LL_{lm}^{l''m''}(r;k) \GF_{l''m''}^{l'm'}(r,r';k) =\one \frac{\delta(r-r')}{r^2}\delta_{ll'}\delta_{mm'}\,.
\label{Eq:GF}
\end{equation}

Suppose now that we know the GF of the system with outgoing boundary conditions and for sources within the sphere, i.e. at $r' \leqslant R$, and we want to determine the system response when it is illuminated by an electromagnetic wave with frequency $\omega$. In the exterior, i.e. for $r>R$, the propagating light can be expanded into a superposition of incoming and outgoing spherical electromagnetic waves~\cite{BohrenBook98}, as derived in \App{App:waveequ} (for explicit expressions see \Sec{Sec:IOEW}). As an example, the expansion of a plane electromagnetic wave into the VSHs as well as into incoming and outgoing vector spherical waves is given in \App{App:PlaneWave}. The illumination can thus be described in open space by a superposition of incoming and outgoing waves of given $l$ and $m$.

To make use of the known GF, which does not contain incoming waves, we replace the incoming waves by sources located exactly on the spherical boundary at $r=R$. This is the only location we can use, since for sources at $r>R$, the GF is not known, and for $r<R$ the incoming wave is generally not the same as in free space. We present two alternative derivations of this conversion.

In Appendices \ref{App:Alternative} and \ref{App:Source}, the conversion of an incoming spherical wave into such a source at $r=R$ is done explicitly, and the field inside the sphere up to its boundary at $r=R$ is then expressed in terms of the GF with this source term. The field on the boundary is equated to the field outside expressed in terms of incoming and outgoing spherical waves, determining the link between the GF and the S-matrix.

Here, instead, we realize the conversion of the illumination into a source on the surface of the sphere by splitting the problem into two parts. In the first part, which contains the given incoming waves in open space, we introduce a surface current at $r=R$, which results in zero electric field inside the system. This part can be solved using only incoming and outgoing waves in open space, disregarding the complexity inside the system, which is ``screened'' from the illumination by the surface current. The second part contains the reverse surface current, no illumination, and a field which can be found using the GF. The sum of both parts then has the given incoming waves and no surface currents, and thus provides the required solution for the illuminated system.

To implement this idea, we split the electric field $\E_{lm}(r)$ into two terms:
\begin{equation}
\E_{lm}(r) = \iE_{lm}(r) \Theta(r-R) + \E^\mathrm{G}_{lm}(r),\label{Eq:EG_iE}
\end{equation}
where the first term, $\iE_{lm}(r)\Theta(r-R)$, is the solution of the first part, and the second term, $\E^\mathrm{G}_{lm}(r)$, is the solution of the second part. The surface current is created by the first derivative of the Heaviside step function $\Theta(r-R)$ in the first term when applying the Helmholtz operator \Eq{Eq:L}, yielding the Dirac delta function $\delta(r-R)$. As shown in \App{App:SurfCurr}, requiring $\iE_{lm}(r)$ to vanish on the surface,
\begin{equation}
[\iE_{lm}(R)]_1=[\iE_{lm}(R)]_2=0,\label{Eq:Econd}
\end{equation}
determines the first part completely, including its outgoing waves, and leads to the surface current
\begin{equation}
\J_{lm}=R
\begin{pmatrix}
R[\iE'_{lm}(R)]_1\\
R[\iE'_{lm}(R)]_2-\sqrt{l(l+1)}[\iE_{lm}(R)]_3\\
0
\end{pmatrix} ,\label{Eq:J}
\end{equation}
so that the second part $\E^\mathrm{G}_{lm}(r)$ becomes the solution of the inhomogeneous Maxwell wave equation
\begin{equation}
\sum_{l'm'}  \LL_{lm}^{l'm'}(r;k) \E^\mathrm{G}_{l'm'}(r)= -\J_{lm}\frac{\delta(r-R)}{r^2}\,.
\label{Eq:EG_2}
\end{equation}
Now, using \Eq{Eq:GF} we find
\begin{equation}
\E^\mathrm{G}_{lm}(r)=-\sum_{l'm'}\GF_{lm}^{l'm'}(r,R;k)\J_{l'm'}\,.\label{EGJ}
\end{equation}

Note that only the second part of the system described by $\E^\mathrm{G}_{lm}(r)$ can mix different $l,m$, which occurs in scattering, while in the first part of the system the incoming and outgoing waves in $\iE_{lm}(r)\Theta(r-R)$ have always the same $l,m$, since this part does not ``see'' what is inside of the system, but only a perfectly conducting sphere. The full solution of the illumination problem is then provided by the superposition of VSHs \Eq{mapping} with $\E_{lm}(r)$ given by \Eq{Eq:EG_iE}.

\section{The S-matrix}\label{Sec:IOEW}
In the empty space outside of the system, the permittivity in VSH representation is $\eps_{lm}^{l'm'}(r)=\one\delta_{ll'}\delta_{mm'}$, and the solutions of the wave equation (\ref{ME-mapped}) consist of two groups: incoming ($d=\mathrm{in}$) and outgoing ($d=\mathrm{out}$) spherical electromagnetic waves, each of which can be split into transverse electric ($p={\rm TE}$) and transverse magnetic ($p={\rm TM}$) polarizations. As derived in \App{App:waveequ}, the solutions have the following form:
\begin{gather}
\E^d_{l,\mathrm{TE}}(r, k)=
\begin{pmatrix}
\tilde{h}_{ld}\\
0\\
0
\end{pmatrix},\label{Eq:TE}\\
\E^d_{l,\mathrm{TM}}(r, k)=\frac{R}{r}
\begin{pmatrix}
0\\
\tilde{\xi}_{ld}\\
\sqrt{l(l+1)} \gamma_{ld}\tilde{h}_{ld}
\end{pmatrix}.\label{Eq:TM}
\end{gather}
Here, we introduced the functions $\tilde{h}_{ld}(r,k)$ and $\tilde{\xi}_{ld}(r,k)$, as well as the coefficients $\gamma_{ld}(k)$, defined as
\begin{gather}
\tilde{h}_{ld}(r,k)= \frac{h_{ld}(kr)}{h_{ld}(kR)}, \quad
\tilde{\xi}_{ld}(r,k)= \frac{\xi'_{ld}(kr)}{\xi'_{ld}(kR)}, \label{Eq:tilde_h}\\
\gamma_{ld}(k) = \frac{h_{ld}(kR)}{\xi'_{ld}(kR)},\label{Eq:gamma}
\end{gather}
where $h_{l,\mathrm{out}}(x)=h^{(1)}_l(x)$ and $h_{l,\mathrm{in}}(x)=h^{(2)}_l(x)$ are the spherical Hankel function of the first and second kind, respectively, $\xi_{ld}(x) = x h_{ld}(x)$ is the Riccati-Bessel function~\cite{Abramowitz1964}, and the prime indicates the derivative.
Note that Eqs.\,(\ref{Eq:TE})--(\ref{Eq:gamma}) are normalized in such a way that on the sphere surface $r=R$ their dominant VSH component is unity,
\begin{equation}
[\E^d_{l,\mathrm{TE}}(R, k)]_1=[\E^d_{l,\mathrm{TM}}(R, k)]_2=1\,.
\label{Eq:Unitary}
\end{equation}

For any real frequency and for $r\geqslant R$, i.e. in homogeneous space,  the electric field can be expressed as a superposition of  incoming and outgoing spherical waves of TE and TM polarization:
\begin{equation}
\E_{lm}(r,k) = \sum\limits_p \left[
A^\mathrm{in}_{lmp} \E^{\mathrm{in}}_{lp}(r,k) +
A^\mathrm{out}_{lmp}\E^{\mathrm{out}}_{lp}(r,k)
\right]\,,\label{Eq:EAEY}
\end{equation}
where $A^\mathrm{in}_{lmp}$ and $A^\mathrm{out}_{lmp}$ are, respectively, the incoming and outgoing amplitudes.
The scattering matrix $S^{l'm'p'}_{lmp}$, which  is defined by
\begin{equation}
A^\mathrm{out}_{lmp} = \sum\limits_{l'm'p'} S^{l'm'p'}_{lmp}A^\mathrm{in}_{l'm'p'}\,,\label{SmatrixDef}
\end{equation}
links the incoming and outgoing amplitudes.

In order to satisfy \Eq{Eq:Econd}, the electric field $\iE_{lm}(r)$ must be chosen as
\begin{equation}
\iE_{lm}(r) = \sum\limits_p A^\mathrm{in}_{lmp}[\E^{\mathrm{in}}_{lp}(r)- \E^{\mathrm{out}}_{lp}(r)]\,,
\label{Eq:EAE}
\end{equation}
considering the normalization  \Eq{Eq:Unitary}. Substituting \Eqs{Eq:TE}{Eq:TM} into \Eq{Eq:EAE}, the electric field $\iE_{lm}(r)$ into \Eq{Eq:J}, and the result for $\J_{lm}$ into \Eq{EGJ}, we find $\E^\mathrm{G}_{lm}(r)$.
Then, equating the fields given by \Eq{Eq:EG_iE} and \Eq{Eq:EAEY} at $r=R$, we obtain a relation between the incoming and outgoing amplitudes, expressed in terms of the GF. Comparing this relation with \Eq{SmatrixDef}, we find the scattering matrix
\begin{equation}
S^{l'm'p'}_{lmp} = {\cal G}^{l'm'p'}_{lmp}(R,R;k) \sigma_{l'p'}-
\delta_{pp'}\delta_{ll'}\delta_{mm'}, \label{SmatrixFinal}
\end{equation}
where
\begin{gather}
\sigma_{l,\mathrm{TE}}(k)=
R\left[\gamma_{l,{\rm out}}^{-1}(k)-\gamma_{l,{\rm in}}^{-1}(k)\right], \label{Eq:J_TE}\\
\sigma_{l,\mathrm{TM}}(k)=
k^2R^3 \left[\gamma_{l,{\rm in}}(k)-\gamma_{l,{\rm out}}(k)\right], \label{Eq:J_TM}\\
{\cal G}^{l'm'p'}_{lmp}(R,R;k) = [\GF^{l'm'}_{lm}(R,R;k)]_{pp'} \, .
\end{gather}

An alternative derivation of \Eq{SmatrixFinal} is presented in \App{App:Alternative}. The derivation is based on a direct comparison of the field inside the sphere, calculated using the GF, and the field outside it, found from the S-matrix. Inside the sphere, the electric field is the solution of Maxwell's wave equation with a delta source term at $r=R$ exactly representing the effect of an incoming spherical wave on the interior region of the sphere. Owing to this delta source, the full electric field within the sphere $r\leqslant R$ is given by the GF which in turn can be found using the ML expansion into RSs of the system and applying the RSE. Outside the sphere, the electric field due to the incoming spherical wave is fully determined by the S-matrix, as it is clear from Eqs.\,(\ref{Eq:EAEY}) and (\ref{SmatrixDef}). Requiring the continuity of the tangent component of the electric field across the boundary $r=R$, we obtain in \App{App:Alternative} the link \Eq{SmatrixFinal} between the GF and the S-matrix.

\section{Mittag-Leffler expansion of the Green's function and the resonant-state expansion} \label{sec:RSE}
In this section, we derive the GF $\g(\r,\r';k)$ of the optical system of interest which we refer to as a {\it new system}, in terms of the analytically known RSs $\E_n(\r)$ of a {\it basis system}.
The derivation is build on the ML expansion of GFs~\cite{More1971,Calderon1976,Bang1978}, which for optical systems results in the following spectral representation of the GF
\begin{equation}
\g(\r,\r';k)  = \sum\limits_\nu \frac{\e_\nu(\r) \otimes \e_\nu(\r')}{2k(k-\varkappa_\nu)}\,,
\label{Eq:g_exp}
\end{equation}
as shown in~\cite{MuljarovEPL10,DoostPRA13,DoostPRA14}. Here, $\otimes$ denotes the direct (dyadic) product.
We assume that the ML expansion \Eq{Eq:g_exp} is valid within the minimal sphere of radius $R$ containing the system.
The electric-field eigenfunctions $\e_\nu(\r)$ with the corresponding wave numbers $\varkappa_\nu$ satisfy Maxwell's wave equation
\be
\LL(\r;\varkappa_\nu)\e_\nu(\r)=0
\ee
and outgoing boundary conditions. Since the GF in \Eq{SmatrixFinal} is expanded into VSHs and projected onto waves of different polarization (TE or TM), it is natural to choose a basis system of spherical symmetry.
The simplest choice of such a system is a homogeneous sphere in vacuum, having a constant permittivity $\epsilon$ different from the one of vacuum, and the radius $R$. This system has known analytic solutions, providing its RSs wave function $\E_n(\r)$ and the eigen wave numbers $k_n$, see e.g. \cite{DoostPRA14}.
The RS wave functions $\e_\nu(\r)$ of the new system are then expanded into the RSs of the basis system,
\be
\e_\nu(\r)=\sum_n C_{n\nu}\E_n(\r)\,.
\label{Exp}
\ee
This expansion is valid for all points inside the sphere $|\r|\leqslant R$.
The GF of the basis system $\G(\r,\r';k)$ has a spectral representation analog of \Eq{Eq:g_exp},
\begin{equation}
\G(\r,\r';k)  = \sum_n \frac{\E_n(\r) \otimes \E_n(\r')}{2k(k-k_n)}\,,
\label{Eq:G_exp}
\end{equation}
which is also valid within the sphere. The RS wave functions $\E_n(\r)$ satisfy the normalization condition~\cite{MuljarovEPL10,DoostPRA14}
\begin{align}
1=&\ \int_{{\cal V}_R}  \varepsilon_b(\br)\En^2\, d\br
\label{norm3}\\
&
+\frac{1}{2 k^2_n}\oint_{S_R} \left[\En\cdot\frac{\partial}{\partial r}r\frac{\partial\En}{\partial r}-r\left(\frac{\partial \En}{\partial r}\right)^2\right] dS\,,
\nonumber
\end{align}
for non-static ($k_n\neq0$) RSs, where ${\cal V}_R$ is the volume and $S_R$ is the surface area of the sphere, and
\be
2=\ \int  \varepsilon_b(\br)\En^2\, d\br
\label{norm4}
\ee
for static ($k_n=0$) RSs, where the integral is extended to the full space. Here, $\varepsilon_b(\br)$ is the permittivity of the basis system, which is $\epsilon$ for $r\leqslant R$ and 1 elsewhere.

The GF $\g(\r,\r';k)$ of the new system, described by the permittivity tensor $\eps(\r)$,
is related to the GF $\G(\r,\r';k)$ of the basis system via the Dyson equation
\begin{multline}
\g(\r,\r';k)  = \G(\r,\r';k)  \\
-k^2 \int_{{\cal V}_R} \G(\r,\r'';k) \Delta \eps(\r'') \g(\r'',\r';k)d\r'', \label{Eq:gG}
\end{multline}
where $\Delta \eps(\r) = \eps(\r) - \one\varepsilon_b(\br)$ is the perturbation. Substituting the expansions \Eq{Eq:g_exp}, \Eq{Exp}, and \Eq{Eq:G_exp} into \Eq{Eq:gG}, we obtain
\begin{multline}
\sum_{nn'}\left( \sum_\nu\left[k \sum_{n''} M_{nn''}C_{n''\nu}-k_n C_{n\nu}\right]\frac{ C_{n'\nu}}{k-\varkappa_\nu}  -
\delta_{nn'} \right) \\ \times\frac{\E_n(\r) \otimes\E_{n'}(\r')}{2k(k-k_n)} = 0, \label{Eq:GG}
\end{multline}
where
\begin{equation}
M_{nn'} = \delta_{nn'}+\frac{1}{2}V_{nn'}\,, \label{Eq:Mdefinition}
\end{equation}
and
\begin{equation}
V_{nn'} = \int_{{\cal V}_R} \E_{n}(\r) \Delta\eps(\r) \E_{n'}(\r)d\r \,. \label{Eq:Vnm}
\end{equation}
are the matrix elements of the perturbation.
To satisfy \Eq{Eq:GG}, it is sufficient to require that
\begin{equation}
\sum_\nu\left[k \sum_{n''} M_{nn''} C_{n''\nu}-k_n C_{n\nu}\right]\frac{ C_{n'\nu }}{k-\varkappa_\nu}  =
\delta_{nn'}. \label{Eq:GG_0}
\end{equation}

While the left hand side of \Eq{Eq:GG_0} is a function of complex variable $k$ having simple poles at $k=\varkappa_\nu$, the right hand side shows that this function is a constant. This requires that the residues at all the poles are zero, leading to the linear matrix eigenvalue problem of the RSE
\begin{equation}
k_n C_{n\nu} = \varkappa_\nu \sum_{n'} M_{nn'} C_{n'\nu},
\label{Eq:EigenvalueProblem}
\end{equation}
which was derived in~\cite{MuljarovEPL10}. Equation~(\ref{Eq:EigenvalueProblem}) determines the eigenvalues $\varkappa_\nu$ and eigenvectors $C_{n\nu}$ of the new system.

Now, replacing the term $k_n C_{n\nu}$ in \Eq{Eq:GG_0} by the sum from \Eq{Eq:EigenvalueProblem}, we obtain the orthonormality of eigenvectors,
\begin{equation}
\sum_{n''\nu}M_{nn''}C_{n''\nu} C_{n'\nu}  = \delta_{nn'}\,. \label{Eq:Normalization_2}
\end{equation}
Treating $M_{nn'}$ and $C_{n'\nu}$ as square matrices, we find  that their product,  $\sum_{n'} M_{nn'}C_{n'\nu}$, is also a square matrix, which, according to \Eq{Eq:Normalization_2}, is the transposed inverse of the matrix $C_{n\nu}$, so that the product of the matrices is an identity matrix. Multiplying these matrices in reverse order also results in the identity matrix, hence we obtain the orthonormality condition,
\begin{equation}
\sum_{n n'}M_{nn'}C_{n\nu} C_{n'\nu'}  = \delta_{\nu \nu'}\,, \label{Eq:Normalization_3}
\end{equation}
which is equivalent to \Eq{Eq:Normalization_2}.
Using the RSE equation (\ref{Eq:EigenvalueProblem}) again, the orthonormality  \Eq{Eq:Normalization_3} can be simplified to
\begin{equation}
\sum_{n} k_n C_{n\nu} C_{n\nu'}  = \varkappa_\nu \delta_{\nu \nu'}\,. \label{Eq:Normalization_4}
\end{equation}

The generalized matrix eigenvalue problem \Eq{Eq:EigenvalueProblem} of the RSE can be reduced to diagonalizing a complex symmetric matrix, as it was shown in~\cite{MuljarovEPL10,DoostPRA14}. Indeed, by making a transformation $C_{n\nu}= \tilde{C}_{n\nu}\sqrt{\varkappa_\nu/k_n}$, the RSE equation (\ref{Eq:EigenvalueProblem}) becomes a symmetric eigenvalue problem
\be
\sum_{n'} \tilde{M}_{nn'} \tilde{C}_{n'\nu}=\lambda_\nu \tilde{C}_{n\nu}
\ee
with the complex symmetric matrix
\be
\tilde{M}_{nn'}=\frac{\delta_{nn'}}{{k_n}} +\frac{V_{nn'}}{2\sqrt{k_n}\sqrt{k_{n'}}}
\label{RSE}
\ee
and the eigenvalues $\lambda_\nu=1/\varkappa_\nu$. Naturally, the orthonormality conditions \Eqs{Eq:Normalization_2}{Eq:Normalization_3}
are also simplified to
\be
\sum_{\nu}  \tilde{C}_{n\nu}\tilde{C}_{n'\nu} = \delta_{n n'}\,,\ \ \  \sum_n  \tilde{C}_{n\nu}\tilde{C}_{n\nu'} = \delta_{\nu \nu'}\,,
\label{ortho}
\ee
as expected for the standard eigenvalue problem. Note that using static ($k_n=0$) modes makes \Eq{RSE} ill-defined. For numerical calculations, however, one can use small finite values of $k_n$ for static modes, with negligible impact on accuracy, while allowing to use the symmetric eigenvalue problem, which can be solved 2-3 times faster \cite{DoostPRA14} than the generalized eigenvalue problem \Eq{Eq:EigenvalueProblem}.

To summarize, the GF $\g(\r,\r';k)$ of the new system is found in the following way. We first find the RSs of the basis system, such as a homogeneous sphere in vacuum, calculating analytically their wave numbers $k_n$ and the wave functions $\E_n(\r)$.  The basis system must be chosen in such a way that the new system is included in its volume.
Then we calculate the matrix elements \Eq{Eq:Vnm} of the difference in the permittivity $\Delta \eps(\r)$ between the new and the basis systems. Then we solve the generalized matrix eigenvalue problem \Eq{Eq:EigenvalueProblem} of the RSE, finding the wave numbers $\varkappa_\nu$ of the new RSs, and the expansion coefficients $C_{n\nu}$, which we normalize according to \Eq{Eq:Normalization_3}. Finally, we use the normalized coefficients $C_{n\nu}$ in the expansion \Eq{Exp}, along with the analytic wave functions $\E_n(\r)$, and then substitute the wave functions $\e_\nu(\r)$ found in this way into the ML expansion \Eq{Eq:g_exp} of the GF.

\section{Results}\label{Sec:Results}
\begin{figure}
	\centering
	\includegraphics[width=\columnwidth]{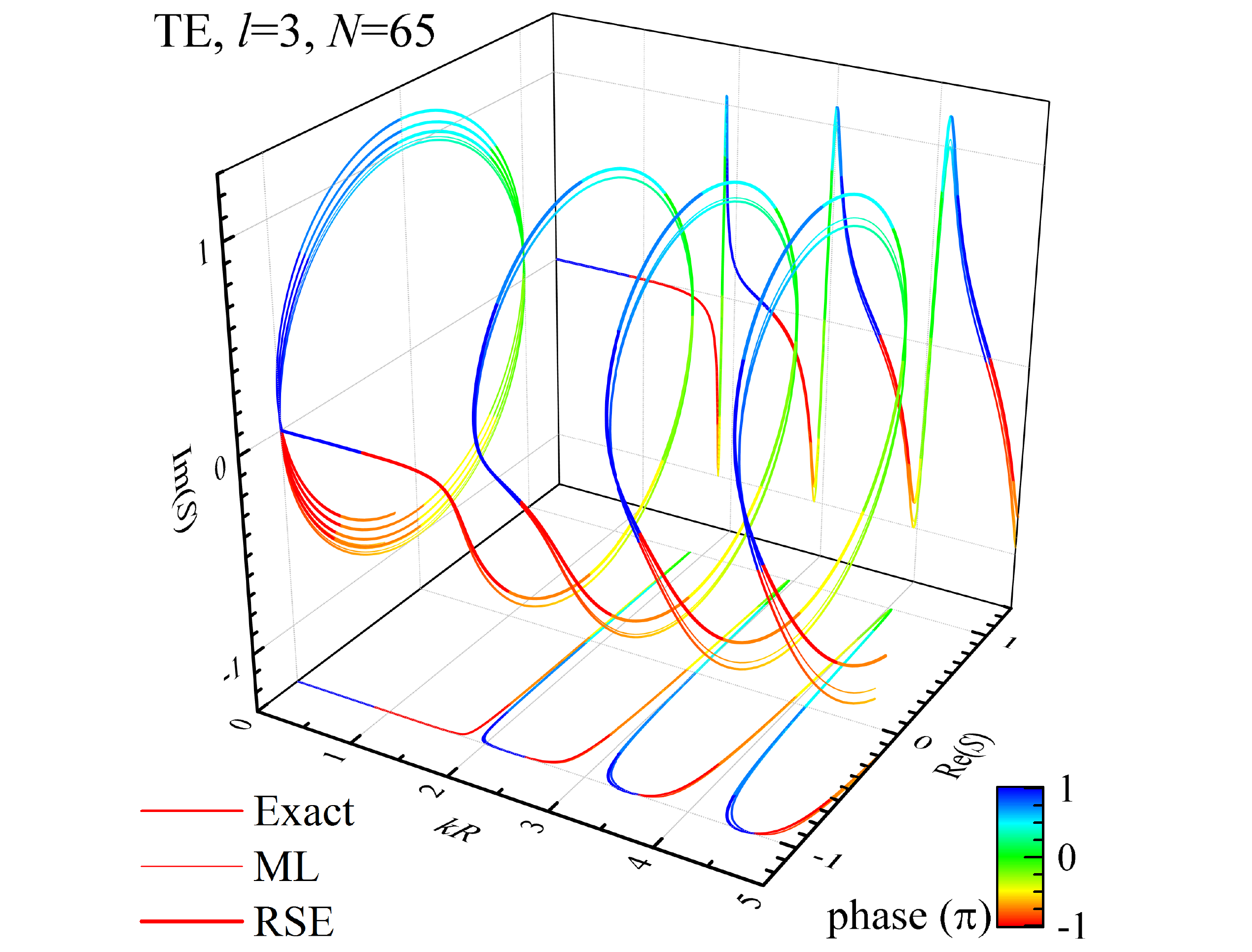}
	\caption{Complex reflection coefficient $S^{lmp}_{lmp}(k)$ for TE polarization ($p={\rm TE}$) and $l=3$ for a dielectric sphere of $\varepsilon=9$ in vacuum. The exact analytic result (normal line) is compared with the ML expansion using the exact RSs for $\varepsilon=9$ (thin line) and with the RSE calculation using a basis sphere of $\epsilon=4$ (thick line), both computed for a basis size of $N=65$. 3D representation with the phase given by the line color. }\label{fig:RefTEl3Plot3D}
\end{figure}

The link \Eq{SmatrixFinal} between the $S$-matrix and the GF, and the ML expansion \Eq{Eq:g_exp} of the GF, in combination with the RSE using a spherically symmetric system as basis, provides a rigorous method of calculating the $S$-matrix of an arbitrary 3D open optical system.
\begin{figure}
	\centering
	\includegraphics[width=\columnwidth]{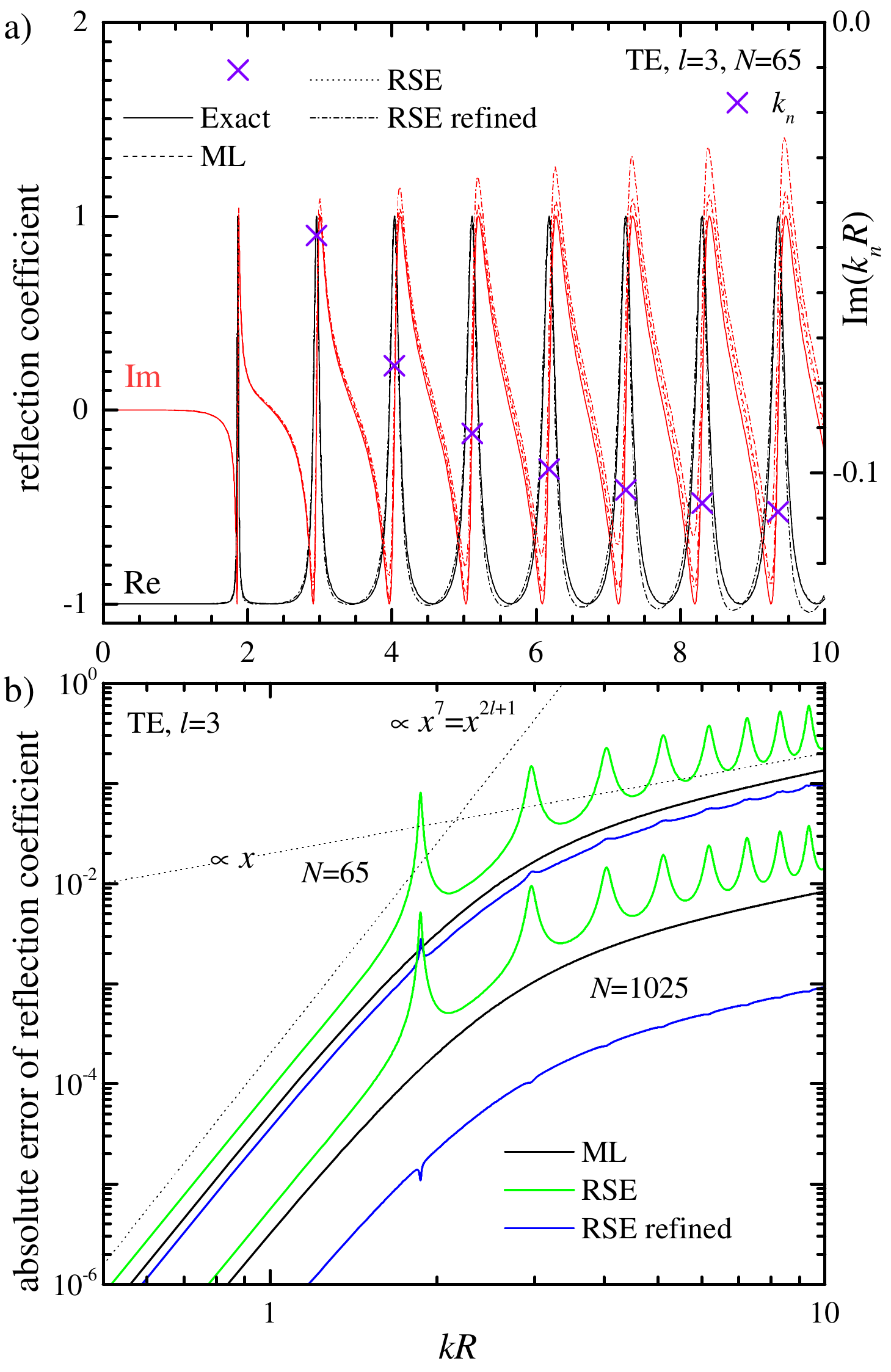}
	\caption{ a) Complex reflection coefficient $S^{lmp}_{lmp}(k)$ for TE polarization ($p={\rm TE}$) and $l=3$  for a dielectric sphere of $\varepsilon=9$ in vacuum. $\Re(S)$ in black, $\Im(S)$ in red. The analytical result (solid line) is compared with the ML expansion using the exact RSs for $\varepsilon=9$ (dashed line) and with the RSE calculation using a basis sphere of $\epsilon=4$ (dash-dotted line), both calculated for a RS basis size of $N=65$, and a refined RSE calculation including a first order correction of the RS fields (dotted line). The RS wavenumbers $k_n$ are given as crosses. b) Absolute error $|S-S^{\rm (exact)}|$ of the ML (black solid line), the RSE (green solid line), and the refined RSE results (blue line). Results for basis sizes $N=65$ and 1025 as labeled. The dotted lines show asymptotic power-law dependencies as labeled.}\label{fig:RefTEl3Plot}
\end{figure}
\begin{figure}
	\centering
	\includegraphics[width=\columnwidth]{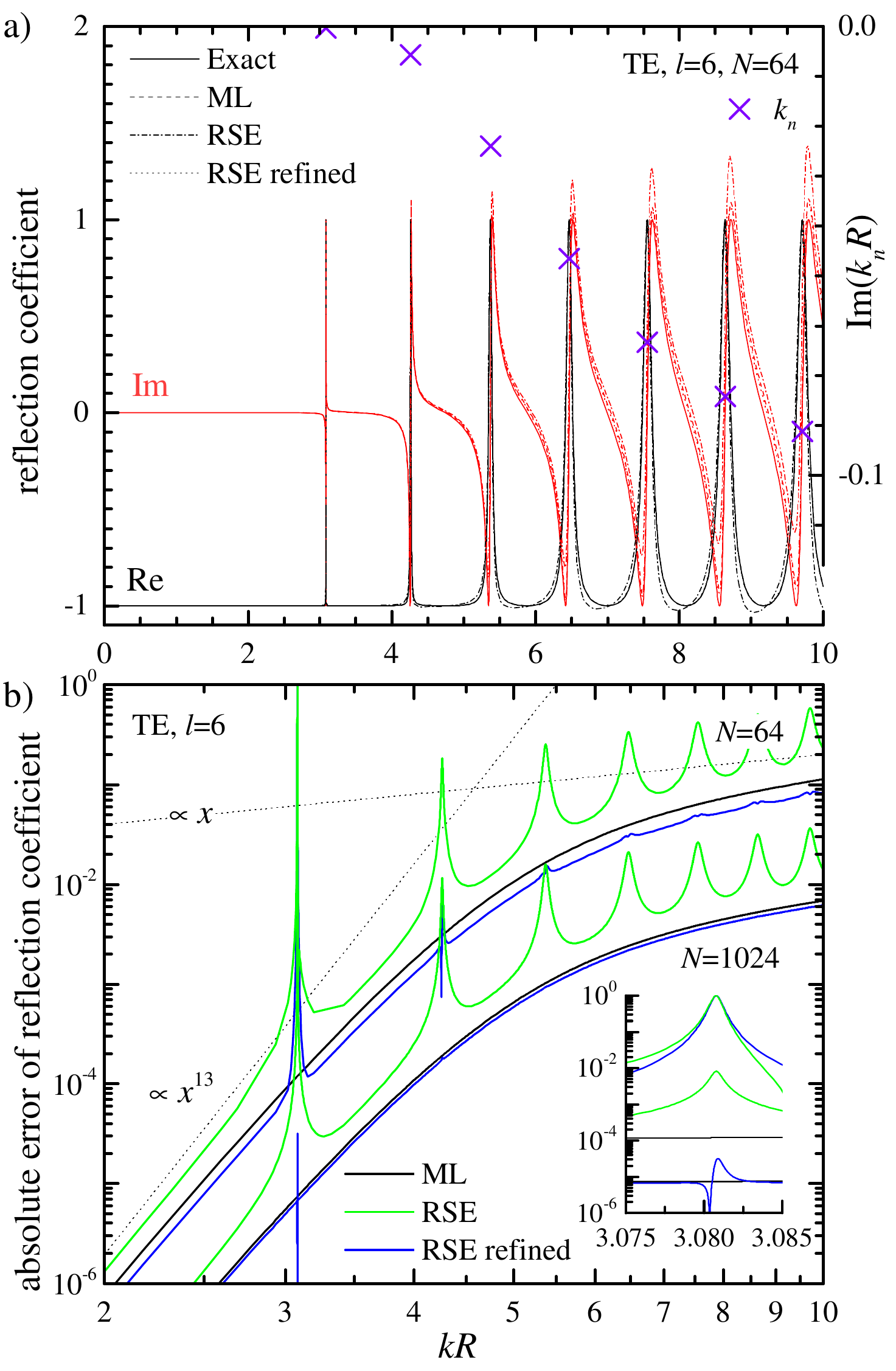}
	\caption{As \Fig{fig:RefTEl3Plot}, but for $l=6$. The inset in b) is a zoom around the fundamental WG RS.}\label{fig:RefTEl6Plot}
\end{figure}

To demonstrate the validity and convergence of this new method, we consider a system which has an exact solution, i.e. a  spherically symmetric system. In particular, in all illustrations below we take a homogeneous dielectric sphere with $\epsilon=4$ in vacuum as basis system and a sphere with equal radius and $\varepsilon=9$ as new system, for which we calculate, the non-zero diagonal $S$-matrix components $S_{lmp}^{lmp}(k)$ as well as the resulting partial and the total scattering cross-sections. As finite basis, we use all RSs with $|k_n|<\kmax$, with a suitably chosen cut-off $\kmax$. The results of this method are called ``RSE'' results. We compare these with the exact analytic solution $S_{lmp}^{\rm (exact)}(k)$ provided in \App{App:AnalyticalSolution_S} and the cross-sections following from it, which is known as Mie theory~\cite{BohrenBook98}.

In addition to this, we use the exact analytic eigenmodes $\e_\nu^{\rm (exact)}$ of the new system in the ML expansion \Eq{Eq:g_exp}, to calculate the corresponding $S$-matrix components  and the scattering  cross-sections. The results of this method are called below ``ML'' results. Technically, this method is equivalent to using a dielectric sphere with permittivity $\epsilon=9$ as basis structure in the RSE.

\subsection{Reflection coefficients}
\label{Sec:Reflection}

The complex reflection coefficient $S^{lmp}_{lmp}(k)$ for TE polarization ($p={\rm TE}$) and $l=3$ is given in \Fig{fig:RefTEl3Plot3D}. The analytical solution $S_{lmp}^{\rm (exact)}(k)$ (see \App{App:AnalyticalSolution_S})  is compared with the RSE result $S^{lmp}_{lmp}(k)$ and with the ML results, both calculated for a RS basis size of $N=65$. Note that in the case of spherical symmetry, $S^{lmp}_{lmp}(k)$ does not depend on $m$, as can be seen explicitly from \Eq{SmatrixFinal}, in which the diagonal GF ${\cal G}^{lmp}_{lmp}(R,R;k)$ is independent of $m$, as it satisfies \Eq{Eq:GF} with the operator $\LL_{lm}^{lm}(r;k)$ being independent of $m$, see \Eq{Eq:L}. Equally,  the analytic solution is also independent of $m$.

We can observe the typical rotation in complex space across the resonances which get broader in $kR$ with increasing $kR$ as they change from whispering gallery (WG) to Fabry-P\'erot (FP) character, which can be traced in the increasing imaginary part of the RS wavenumbers $k_n$ shown in \Fig{fig:RefTEl3Plot}a. Since the sphere is non-absorbing, the analytical reflection coefficient has unity magnitude, as is clearly visible in the $(\Re(S),\Im(S))$ projection. Both expansions of the GF are deviating from the analytic result in a similar manner, with the error being dominated by a positive shift in $\Im(S)$. In the RSE results, which are using approximate RSs, this deviation is a few times larger than in the ML results which are using the exact RSs instead.

In \Fig{fig:RefTEl3Plot}a the data given in \Fig{fig:RefTEl3Plot3D} is shown separated into real and imaginary parts. \Fig{fig:RefTEl3Plot}b shows the absolute error $|S-S^{\rm (exact)}|$ of the RSE and  ML results for different basis sizes $N$. We can observe that the error is scaling as $N^{-1}$. Furthermore, the error scales as $k$ for $kR \gg l$, and as $k^{2l+1}$ for $kR \ll l$. The former is consistent with the above mentioned $N^{-1}$ scaling, since the error scales as $k/k_{\rm max}$, with $k_{\rm max}\propto N$. The latter is due to the asymptotic behaviour of the illumination field.
\begin{figure}
	\centering
	\includegraphics[width=\columnwidth]{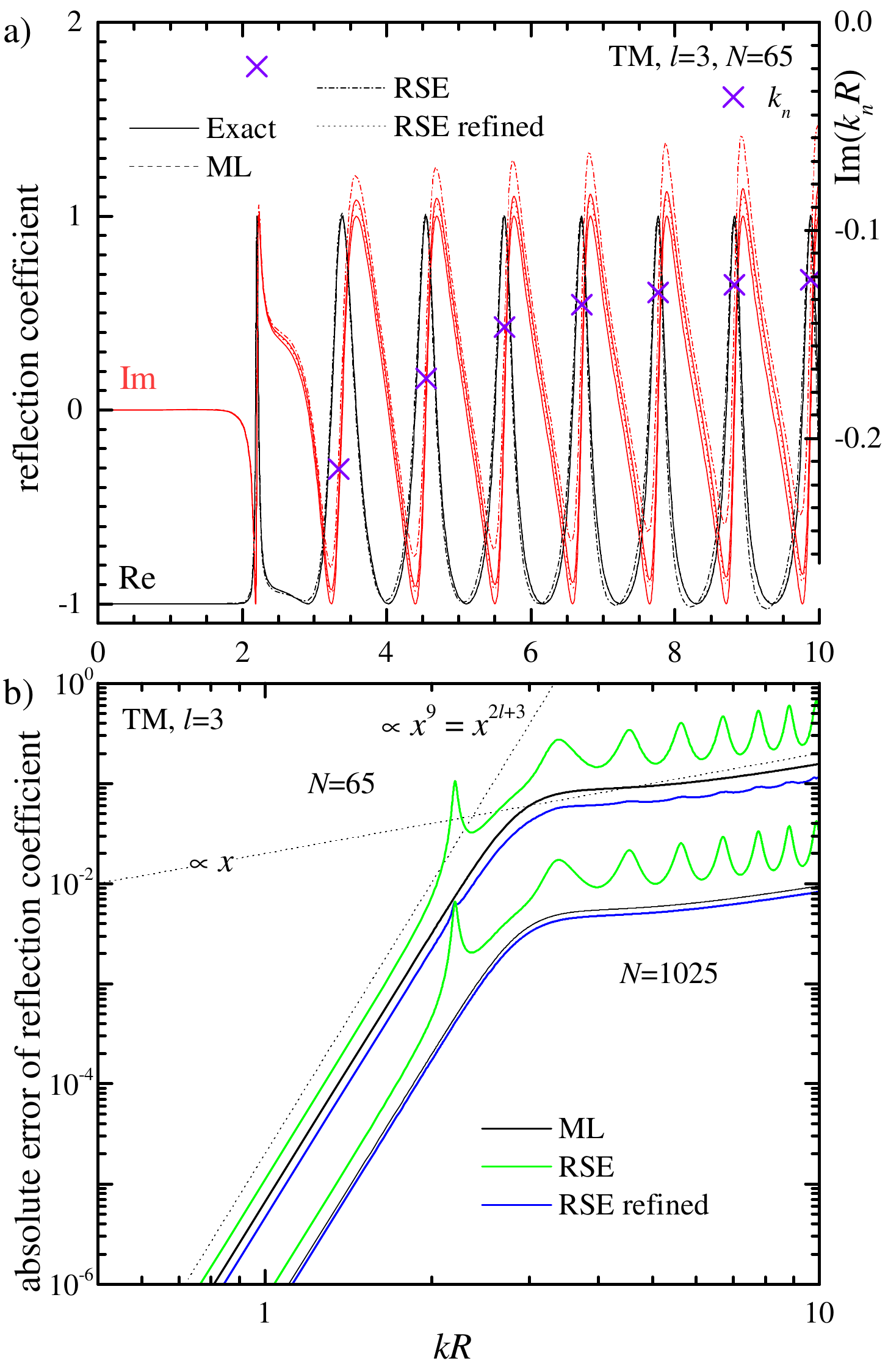}
	\caption{As \Fig{fig:RefTEl3Plot}, but for TM polarization ($p={\rm TM}$).}\label{fig:RefTMl3Plot}
\end{figure}
\begin{figure}
	\centering
	\includegraphics[width=\columnwidth]{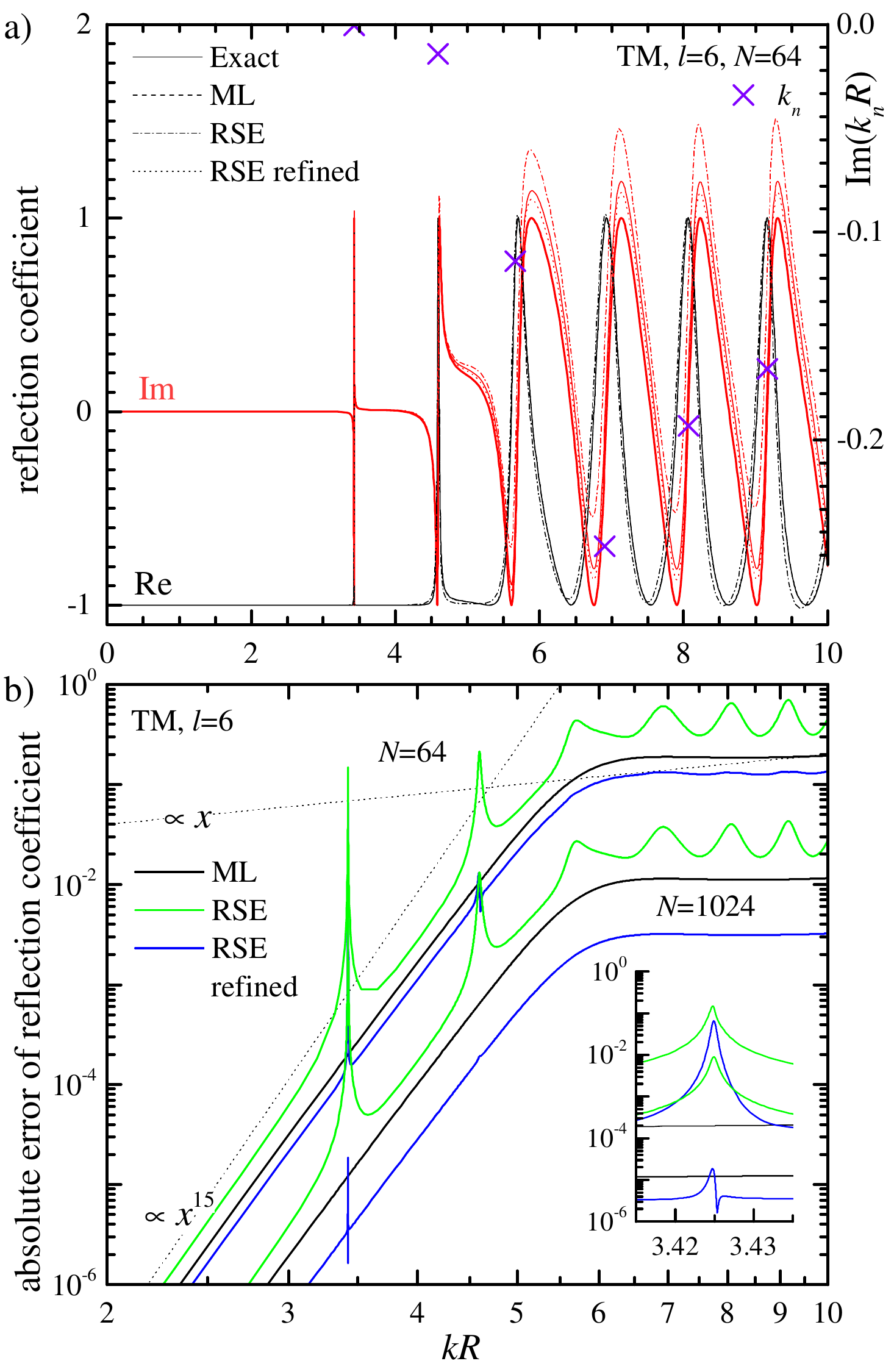}
	\caption{As \Fig{fig:RefTEl6Plot}, but for TM polarization.($p={\rm TM}$)}\label{fig:RefTMl6Plot}
\end{figure}

Notably, the error of the ML results does not show any resonant features, since its error is due to the missing high-frequency RSs.
The error of the RSE calculation instead shows peaks at the resonances, scaling as $N^{-1}$. This could be due to either the error in the RS frequencies $\varkappa_\nu$, or due to the error in the RS fields $\e_\nu$. However, the error in $\varkappa_\nu$ scales as $N^{-3}$ \cite{DoostPRA14}, which is inconsistent with the observed scaling. We therefore conclude that the error is given by the error in the fields $\e_\nu$, as has been observed already in the one-dimensional case for which the scattering problem is simpler to treat \cite{DoostPRA12}.
\begin{figure}
	\centering
	\includegraphics[width=\columnwidth]{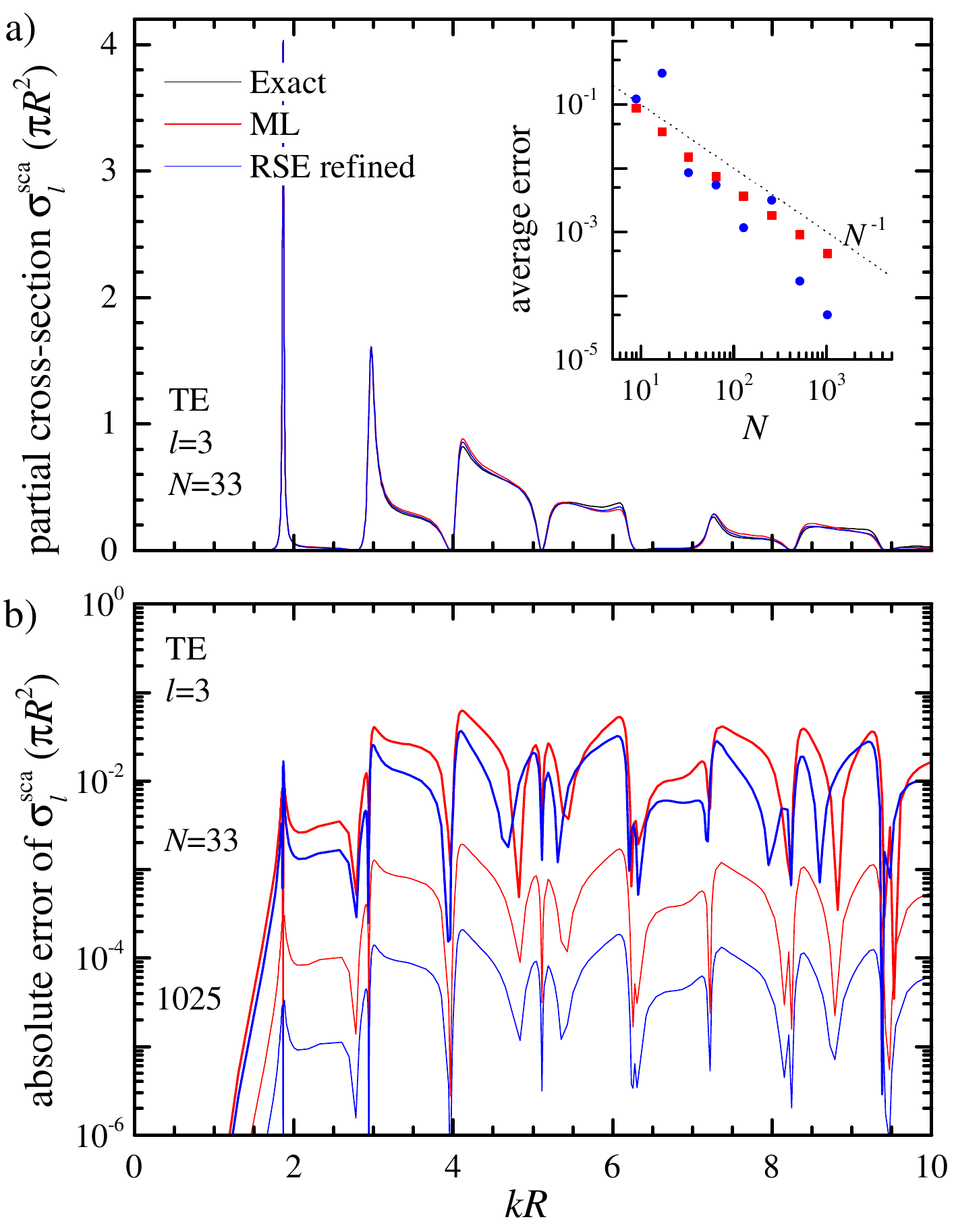}
	\caption{ a) Partial scattering cross-section $\sigma^{\rm sca}_{l}(k)$ for TE polarization and $l=3$ for a dielectric sphere of $\varepsilon=9$ in vacuum. The analytical result (black) is compared with the ML expansion for a basis size of $N=33$, using the exact RSs for $\varepsilon=9$ (red line), and the RSs determined by the RSE with refinement, using as basis a sphere with $\epsilon=4$ (blue line). b) Absolute error $|\sigma^{\rm sca}_{l}(k)- \sigma^{\rm sca (exact)}_{l}(k)|$ for ML (red line) and RSE calculation (blue line). Results for a basis size $N=33$ are given as thick lines, and for $N=1025$ as thin lines. Inset in a): absolute error averaged over the range of $0\leqslant kR\leqslant 10$, as function of the basis size $N$. The dotted line gives $N^{-1}$.}\label{fig:PSTEl3Plot}
\end{figure}
\begin{figure}
	\centering
	\includegraphics[width=\columnwidth]{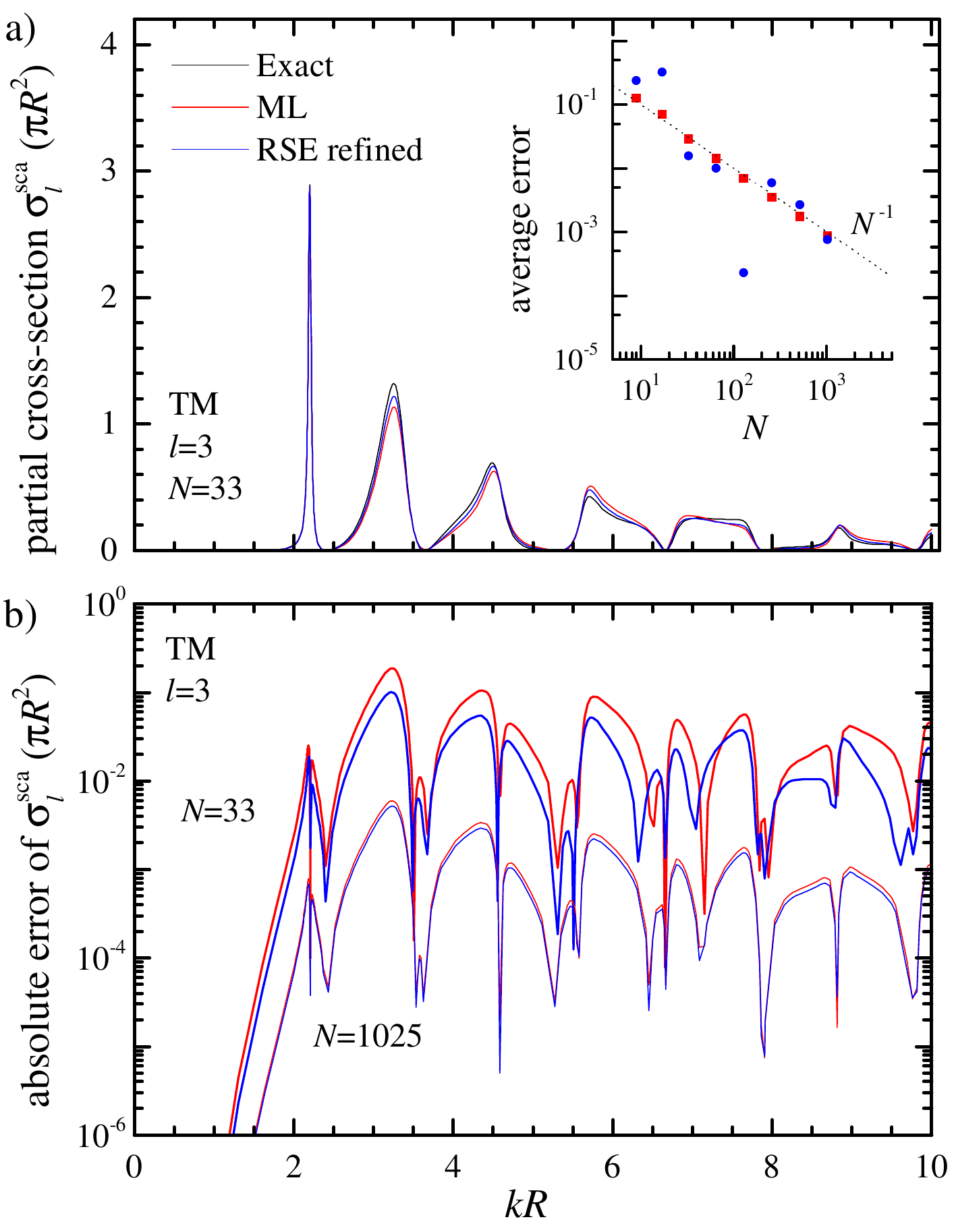}
	\caption{ As \Fig{fig:PSTEl3Plot} but for TM polarization ($p={\rm TM}$).}\label{fig:PSTMl3Plot}
\end{figure}

To improve the error of the fields, we have added a first-order correction of the fields, extending the basis size from $N$ to $N+L$ and treating the additional $L$ modes in first perturbation order, as described in~\App{App:TruncatedSpace}. The size of the basis is chosen such that $L\approx N^2$. Using this correction, the error due to the fields becomes insignificant, and the remaining error is dominated by the missing high-frequency RSs, both in the RSE and the ML results. Interestingly, the error of these ``refined'' RSE results can be even lower than the ML results, as is specifically evident for $N=1025$. This indicates that the remaining error in the RSs can partially compensate the contribution of the missing RSs.

Moving to $l=6$, shown in \Fig{fig:RefTEl6Plot}, we find a behavior consistent with the previous discussion. The high angular momentum supports sharper WG RSs, such as the first two modes, at $kR\sim3.1$ and $kR\sim4.3$ (Q-factors are about 4500 and 340, respectively).
For the sharp WG RSs, we find that the error in the RS wave number is relevant for $N=64$, as can be seen in the inset of \Fig{fig:RefTEl6Plot}b. We note that the error of the wave numbers can be improved by 1-2 orders of magnitude by extrapolation \cite{DoostPRA12}.

Switching to TM modes, we show the corresponding results in \Fig{fig:RefTMl3Plot} for $l=3$ and in \Fig{fig:RefTMl6Plot} for $l=6$. The behavior is similar to the TE polarization. However, the asymptotics of the error for $kR \ll l$ is now $k^{2l+3}$, again due to the scaling of the illumination field.
The somewhat different behavior of TE and TM fields around $kR\sim l$ can be understood considering the Fresnel reflection at the sphere surface -- while TE fields correspond to s-polarized light and therefore show a monotonous increase of Fresnel reflection with incident angle, the TM fields correspond to p-polarization and exhibit a reduced reflection around Brewster's angle. The linewidths of the TM RSs in this region, which occurs around $kR=l$, are therefore much larger than for the TE RSs, as can be seen in the data.

\subsection{Partial scattering cross-sections}

The scattering cross-section $\ssc$ can be determined from the S-matrix as detailed in~\App{App:SC}.
The total cross-section is a sum over all partial contributions $\sscl$ from each angular momentum $l$. For a spherically symmetric system, each partial cross-section $\sscl$ can be found using the spectral representation of the GF, as a sum over the RSs for the given $l$. Also, the analytic form of $\sscl$ for a dielectric sphere in vacuum is known~\cite{BohrenBook98}, see \App{App:AnalyticalSolution_S}. The corresponding $\sscl$ for $p={\rm TE}$, $l=3$ and $N=33$ is given in \Fig{fig:PSTEl3Plot}a. The typical spectral structure consisting of sharp WG RSs, which are getting broader with increasing $kR$, is visible. The results for both GF expansions (ML and RSE refined) show a similar error, which is given in \Fig{fig:PSTEl3Plot}b, and compared with a larger basis size $N=1025$.
To investigate the convergence, we show in the inset of \Fig{fig:PSTEl3Plot}a the error averaged over the displayed range of $kR$, for increasing basis size $N$. We find an error scaling as $N^{-1}$, the same as for the reflection coefficients.
Notably, while the scaling is well defined for the ML calculation, the refined RSE results are showing some fluctuations. This indicates that the first-order correction of the fields, while generally sufficient for an error limited by the cut-off of the expansion, contains additional influences, consistent with the discussion concerning the reflection coefficients.
A similar qualitative behaviour is found for the TM modes, as shown in \Fig{fig:PSTMl3Plot}.

\begin{figure}
	\centering
	\includegraphics[width=\columnwidth]{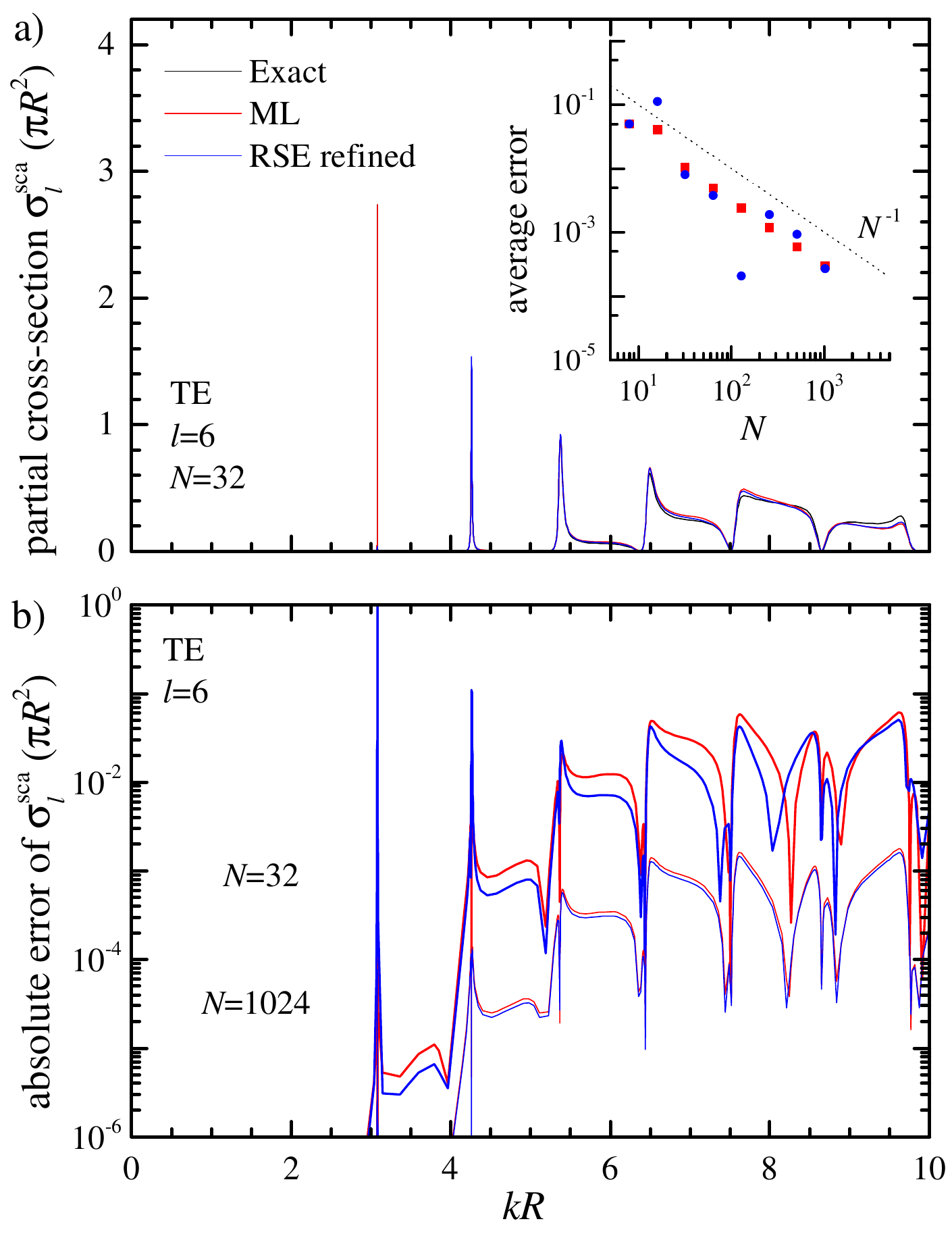}
	\caption{ As \Fig{fig:PSTEl3Plot} but for $l=6$, and basis sizes $N=32$ and $N=1024$ as labelled.}\label{fig:PSTEl6Plot}
\end{figure}
\begin{figure}
	\centering
	\includegraphics[width=\columnwidth]{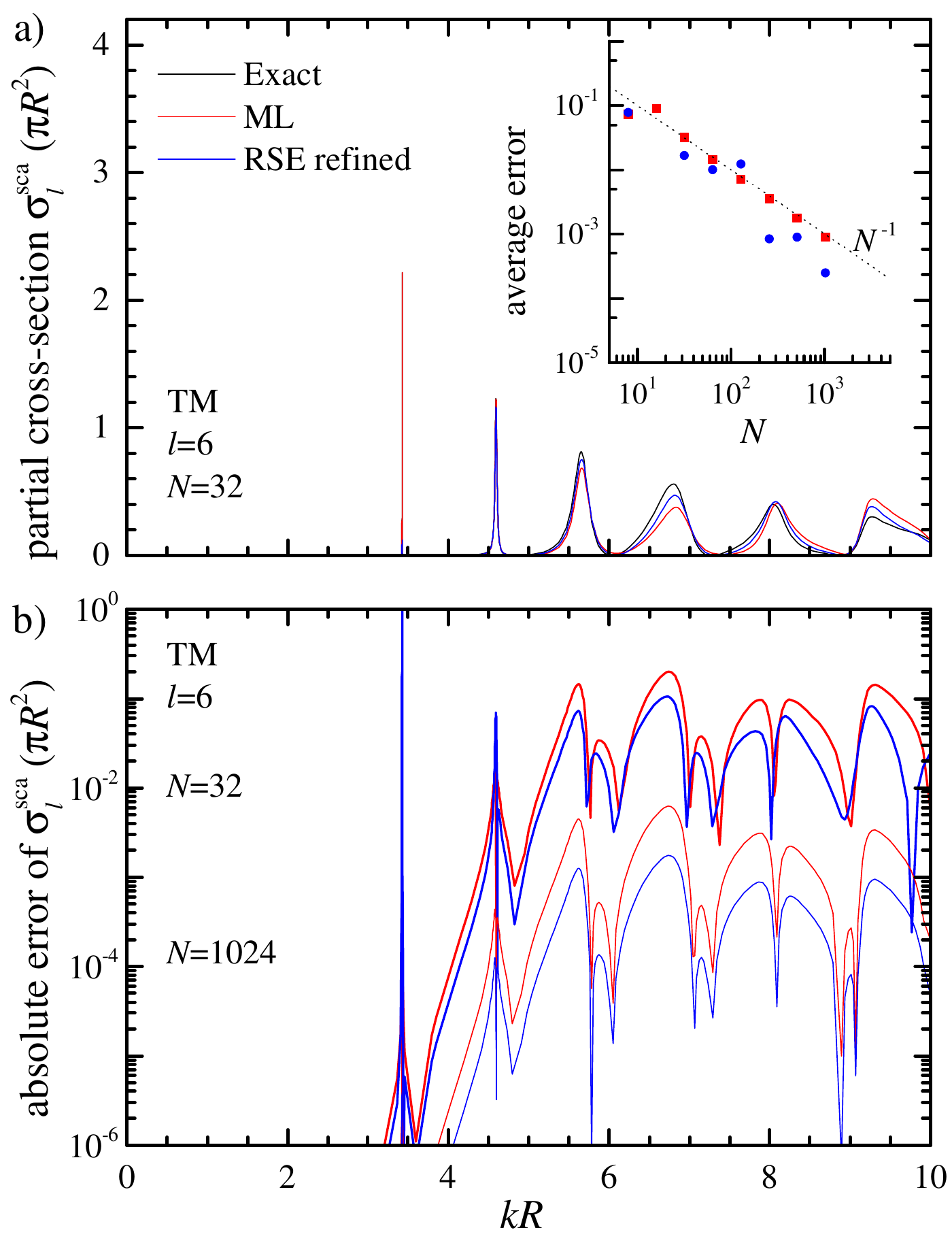}
	\caption{ As \Fig{fig:PSTEl6Plot} but for TM polarization ($p={\rm TM}$).}\label{fig:PSTMl6Plot}
\end{figure}
Moving to TE, $l=6$, shown in \Fig{fig:PSTEl6Plot}, the first WG RS has a rather narrow linewidth, which is smaller than the error in the RS wave numbers for $N=32$. This leads to a large error close to this resonance.
We note that generally, the linewidth of the fundamental WG RS decreases exponentially with increasing $l$, so it quickly becomes less than the RS wave number error, which in turn scales as a power law, namely $N^{-3}$. In typical realistic systems, however, the mode linewidth is limited by absorption and surface roughness, rendering this issue less relevant for practical applications. Again, similar results are observed for the TM modes, as shown in \Fig{fig:PSTMl6Plot}.

\subsection{Total scattering cross-section}

\begin{figure}
	\centering
	\includegraphics[width=\columnwidth]{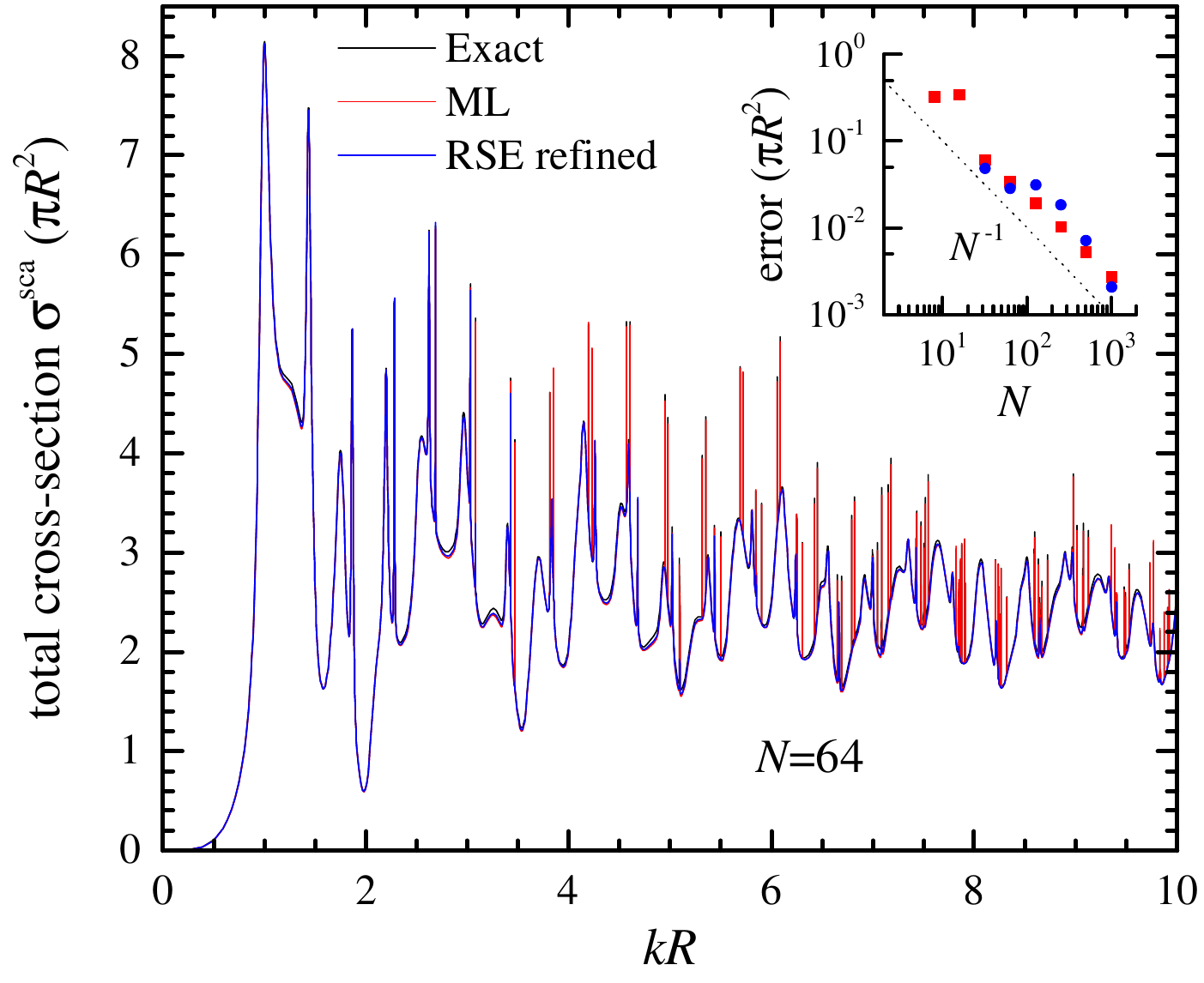}
	\caption{Total scattering cross-section $\sigma^{\rm sca}(k)$ for a dielectric sphere of $\varepsilon=9$ in vacuum. The analytical result using using Mie theory~\cite{BohrenBook98} (black), is compared with ML expansion for a basis size of $N=64$, using the exact RSs for $\varepsilon=9$ (red line), and the RSs determined by the RSE with refinement, using as basis a sphere with $\epsilon=4$  (blue line). Inset: absolute error averaged over $0\leqslant kR\leqslant 10$, as function of the basis size $N$. The dotted line gives $N^{-1}$.}\label{fig:SPlot}
\end{figure}

Summing over all partial scattering cross-sections for $l\leqslant 20$, for a given $N$ used for each $l$, we arrive at the total scattering cross-section presented in \Fig{fig:SPlot}.
The spectrum shows a large number of peaks, having an exponentially decreasing minimum width with increasing $kR$, as discussed above. For $kR \gg 1$, twice the geometrical cross-section is approached, as expected considering diffraction, which is known as the extinction paradox~\cite{BohrenBook98}. In the limit $kR \ll 1$, the well-known scaling in the Rayleigh scattering regime $\ssc\propto R^6$ is seen.

The error averaged over the displayed range of $kR$ is given in the inset versus $N$. We find again an error scaling as $N^{-1}$,
both for the exact RSs and the ones calculated via the RSE, using the refinement. Notably, the remaining error is due to the missing high frequency RSs in the GF expansion. It is conceivable that one can approximate the contribution of these RSs using the high-frequency behaviour of the GF. This will be of relevance for applications of the RSE to scattering and absorption of non-spherical objects, which do not allow using symmetry to factorize the problem and require all RSs within $k<k_{\rm max}$ to be taken into account in the RSE.

We can thus see that the RSE is well suited to determine RSs for scattering calculations, providing errors comparable to those coming from using the exact RSs. The observed $N^{-1}$ convergence of $\ssc$ leads to errors in the 1\% range for about $N=100$ RSs in the  spherically symmetric case treated in the example shown.

\section{Summary and Conclusions}
In summary, we have presented and verified an exact method to determine the scattering matrix of a finite three-dimensional system using its resonant states. This was achieved by expressing incoming and outgoing waves in the basis of vector spherical harmonics, expanding the Green's function into resonant states, and determining the resonant states using the resonant-state expansion with a spherically symmetric basis. The accuracy of the method is limited only by the basis size used, which is determined by the wave-number cut-off. We have demonstrated the convergence of the calculation of the scattering matrix and the scattering cross-section with an error scaling as the inverse basis size. It is conceivable that this convergence can be improved in future work, e.g. by introducing approximate treatments of the states with frequencies above the cut-off, which are not included in the basis.

The presented method establishes a new paradigm, based on the resonant-state expansion, for calculating the response of finite three-dimensional open optical systems, which might have significant practical applications in the future.


\begin{acknowledgments}
This work was supported by the S\^er Cymru National Research Network in Advanced Engineering and Materials, the EPSRC under grant EP/M020479/1, and by RFBR (Grant No. 16-29-03283).
\end{acknowledgments}

\appendix

\section{VSH representation of vectors, tensors and operators}
\label{App:VSH}

We define VSHs according to \Eqs{Eq:Y1}{Eq:Y3}, in which
\begin{equation}
Y_{lm}(\Omega) = \sqrt{\frac{2l+1}{2}\frac{(l-|m|)!}{(l+|m|)!}}P^{|m|}_l(\cos\theta)\chi_m(\varphi)\,,
\label{Eq:Y}
\end{equation}
are the {\it real} spherical harmonics, $P^m_l(x)$ are the associated Legendre polynomials, $l$ ($l=1,2,...$) and $m$ ($-l\leqslant m\leqslant l$) are the spherical quantum numbers,
and the azimuthal functions are defined as
\be \chi_m(\varphi)=\left\{
\begin{array}{lll}
\pi^{-1/2}\sin(m\varphi) & {\rm for} & m<0\, \\
(2\pi)^{-1/2} & {\rm for} & m=0\, \\
\pi^{-1/2}\cos(m\varphi) & {\rm for} & m>0\,,
\end{array}
\right. \label{chi-n} \ee
i.e. in the same way as in~\cite{DoostPRA14}.

The VSHs constitute a complete set of vector functions on a unit sphere. Therefore any vector field can be expanded into VSHs, as given by \Eq{mapping}, in which the radially dependent expansion coefficients are
\be
[\E_{lm}(r)]_i = \int \Y_{ilm}(\Omega)\cdot \E(\r) d\Omega\,.
\label{coef}
\ee
Equations (\ref{mapping}) and (\ref{coef}) determine the mapping of a vector field $\E(\r)$ in real space into a vector field $\E_{lm}(r)$ in the space of VSHs.
Similarly, a tensor $\eps(\r)$ in real space is mapped into a tensor $\eps_{lm}^{l'm'}(r)$ in VSH space,
\be
[\eps_{lm}^{l'm'}(r)]_{ij} = \int \Y_{ilm}(\Omega)\cdot \eps(\r) \Y_{jl'm'}(\Omega)d\Omega\,,
\ee
and a dyadic GF $\GF(\r,\r';k)$ into $\GF_{lm}^{l'm'}(r,r';k)$ given by
\begin{multline}
[\GF_{lm}^{l'm'}(r,r';k)]_{ij}=\\
\int \int\Y_{ilm}(\Omega)\cdot \GF(\r,\r';k) \Y_{jl'm'}(\Omega')d\Omega d\Omega'\,.
\end{multline}
To transform the operator  $\LL(\r)$ in \Eq{ME} to $\LL_{lm}^{l'm'}(r)$ in \Eq{ME-mapped},
we first consider the curl operator. To find its mapping onto the VSH space, we note that it is diagonal in $l$ and $m$. Indeed, 
\bea
r\nabla\times \Y_{1lm}(\Omega)&=& - \Y_{2lm}(\Omega)-\alpha_l \Y_{3lm}(\Omega)\,\nonumber\\
r\nabla\times \Y_{2lm}(\Omega)&=& \Y_{1lm}(\Omega)\,\\
r\nabla\times \Y_{3lm}(\Omega)&=& -\alpha_l \Y_{1lm}(\Omega)\nonumber\,,
\eea
where $\alpha_l=\sqrt{l(l+1)}$, and also
\bea
\r\times \Y_{1lm}(\Omega)&=& - r \Y_{2lm}(\Omega)\,\nonumber\\
\r\times \Y_{2lm}(\Omega)&=& r \Y_{1lm}(\Omega)\,\\
\r\times \Y_{3lm}(\Omega)&=& 0\nonumber\,,
\eea
showing that the operators $\nabla\times$ and $\r\times$ when acting on the VSHs, do not alter their angular quantum numbers. Now, taking the VSH expansion of an arbitrary vector function ${\bf f}(\r)$ with fixed $l$ and $m$,
\be
{\bf f}(\r)=\sum_i f_i(r) \Y_{ilm}(\Omega)\,,
\ee
and using the fact that
\be
\nabla\times f_i(r) \Y_{ilm}(\Omega)= f_i(r)\nabla\times \Y_{ilm}(\Omega)+\frac{f'_i(r)}{r} \r\times \Y_{ilm}(\Omega)\,,
\ee
where the prime indicates the differentiation with respect to $r$, we find
\be
\nabla\times{\bf f}(\r)=\sum_i g_i(r) \Y_{ilm}(\Omega)\,,
\ee
with
\be
g_1=\frac{1}{r}(r f_2)'-\frac{\alpha_l}{r} f_3\,,\ \ \
g_2=-\frac{1}{r}(r f_1)',\ \ \
g_3=-\frac{\alpha_l}{r} f_1\,.
\label{f-to-g}
\ee
Applying the curl operator again, we find the mapping of the double curl operator $\nabla\times\nabla\times \to \hat{L}_{l}(r)$, which is given by
\be
\nabla\times\nabla\times f_i(r) \Y_{ilm}(\Omega) = \sum_j [\hat{L}_{l}(r)]_{ij}  f_j(r) \Y_{jlm}(\Omega)\,.
\ee
The matrix $\hat{L}_{l}(r)$ is found by using \Eq{f-to-g} twice, yielding
\begin{equation}
\hat{L}_{l}(r)=
\begin{pmatrix}
\vspace{1mm}
\displaystyle
-\frac{1}{r}\frac{d^2}{dr^2} r +\frac{\alpha_l^2}{r^{2}} & 0 & 0\\
\vspace{2mm}
0 & \displaystyle -\frac{1}{r}\frac{d^2}{dr^2} r & \displaystyle \frac{\alpha_l}{r}\frac{d}{dr} \\
0 & \displaystyle -\frac{\alpha_l}{r^{2}}\frac{d}{dr} r &\displaystyle\frac{\alpha_l^2}{r^{2}}
\end{pmatrix},\label{Eq:L2}
\end{equation}
and is shown in \Eq{Eq:L} resubstituting $\alpha_l$.

\section{Solutions of the wave equation in free space}
\label{App:waveequ}

In empty space, the operator $\LL_{lm}^{l'm'}(r)$ in the wave equation (\ref{ME-mapped}) becomes
\be
\LL_{lm}^{l'm'}(r) =\left[ -\hat{L}_{l}(r) + k^2 \right] \delta_{ll'}\delta_{mm'}\,,
\label{L-free}
\ee
and thus \Eq{ME-mapped} reduces to
\be
-\hat{L}_{l}(r) \E(r) +k^2 \E(r)=0\,,
\label{wave-equ}
\ee
where $\E(r)$ is the electric field in VSH space,
having components $E_i(r) = [\E(r)]_i$ ($i=1\,,2\,,3$), with indices $l$ and $m$ omitted for brevity.
Due to the block-diagonal form of the $\hat{L}_{l}(r)$ (see \Eq{Eq:L2}), the wave equation (\ref{wave-equ}) splits into TE and TM polarizations.

The TE polarization is given by a single scalar differential equation
\be
\frac{d^2}{dr^2} r E_1 - \frac{\alpha_l^2}{r} E_1+k^2r E_1=0\,,
\ee
which is a spherical Bessel equation for $E_1$, having solutions in a form of spherical Hankel functions,
\be
E_1(r)=\tilde{h}_{ld}(r,k)\,,
\label{E1}
\ee
here normalized in such a way that $E_1(R)=1$, see \Eq{Eq:tilde_h} for the definition of $\tilde{h}_{ld}$. The resulting full electric field in the TE polarization is then given by \Eq{Eq:TE}.

The TM polarization is described by a pair of coupled differential equations following from \Eq{wave-equ}:
\bea
\frac{d^2}{dr^2} r E_2 - \alpha_l \frac{d}{dr}  E_3+k^2r E_2&=&0\,.
\label{TM-equ1}
\\
\frac{d}{dr} r E_2 - \alpha_l E_3+\frac{k^2r^2}{\alpha_l} E_3&=&0\,.
\label{TM-equ2}
\eea
Excluding $E_2$, we again obtain a spherical Bessel equation,
\be
\frac{d^2}{dr^2} r^2 E_3 - \frac{\alpha_l^2}{r} E_3+k^2r^2 E_3=0\,,
\ee
this time for $rE_3$. Choosing the normalization of the field in such a way that $E_2(R)=1$ (see below), we obtain
\be
rE_3(r)=R\alpha_l \gamma_{ld}\tilde{h}_{ld}(r,k)
\label{E3}
\ee
where $\gamma_{ld}$ is defined in \Eq{Eq:gamma}. Finally, combining \Eq{TM-equ1} and \Eq{TM-equ2}, we find
\be
E_2(r)=\frac{1}{\alpha_l r} \frac{d}{dr} r^2 E_3(r)=\frac{R}{r}\tilde{\xi}_{ld}(r,k)
\label{E2}
\ee
where $\tilde{\xi}_{ld}$ is given by \Eq{Eq:tilde_h}, which provides the normalization $E_2(R)=1$ used above. Together with \Eq{E3}, this yields the full electric field in the TM polarization given by \Eq{Eq:TM}.

Note that, according to the chosen normalization, the VSH fields of the TE and TM spherical waves on the surface of the sphere $r=R$ are given by
\be
\E^d_{l,{\rm TE}}(R,k)=\left(\begin{array}{c}
1\\ 0\\0
\end{array}\right),
\ \
\E^d_{l,{\rm TM}}(R,k)=\left(\begin{array}{c}
0\\ 1\\ \alpha_l\gamma_{ld}(k)
\end{array}\right)\,,
\label{norm}
\ee
in agreement with \Eq{Eq:Unitary}

\section{Link between the S-matrix and the Green's function: An alternative derivation of \Eq{SmatrixFinal} }
\label{App:Alternative}

Owing to the linearity of \Eq{ME}, the electric field for an arbitrary excitation of the system is a linear combination of spherical waves.
For a single incoming spherical wave with given spherical quantum numbers $l'$ and $m'$ and polarization $p'$, where either $p'={\rm TE}$ or $p'={\rm TM}$, we have the incoming amplitudes $A^{\rm in}_{lmp}=\delta_{ll'} \delta_{mm'} \delta_{pp'}$ in \Eq{Eq:EAEY}. Using this excitation condition and the $S$-matrix defined by \Eq{SmatrixDef}, we obtain  the full electric field in the VSH representation,
\be
\E_{lm}(r)=\delta_{ll'} \delta_{mm'} \E^{\rm in}_{lp'}(r,k)+\sum_{p} S_{lmp}^{l'm'p'}(k) \E^{\rm out}_{lp}(r,k)\,,
\label{E-SM}
\ee
which is valid for $r\geqslant R$. At the same time, for  $r\leqslant R$, i.e. inside the basis sphere containing the system, the  electric field $\E_{lm}(r)$ can be determined via the RSE using the GF and a suitable excitation source. The GF is expressed as the ML expansion into the RSs of the new system, which are in turn found from the RSs of the basis system, applying the RSE, see \Sec{sec:RSE}. Since the GF satisfies the wave equation  with a $\delta$-source term \Eq{Eq:GF}, it is necessary to represent the effect of the incoming spherical wave on the region $r\leqslant R$ by a spherical $\delta$-source term at $r=R$.
The electric field $\E_{lm}(r)$ for $r\leqslant R$ then becomes the solution of the inhomogeneous Maxwell wave equation
\be
\sum_{l''m''}  \LL_{lm}^{l''m''}(r;k) \E_{l''m''}(r) = \bs_{l'p'} \frac{\delta(r-R)}{R^2}\delta_{ll'}\delta_{mm'}
\label{ME-source}
\ee
with the source vectors  $\bs_{lp}$ given by
\be
\bs_{lp}= \sigma_{lp} {\bf e}_p\,,\ \ \  {\bf e}_{\rm TE}=\left(\begin{array}{c}
1\\ 0\\0\end{array}\right), \ \ \  {\bf e}_{\rm TM}=\left(\begin{array}{c}
0\\ 1\\0
\end{array}\right),
\label{sigmas}
\ee
as derived in \App{App:Source}. Here, $\sigma_{lp}$ are defined by \Eqs{Eq:J_TE}{Eq:J_TM}.
Solving \Eq{ME-source} with the help of the GF satisfying \Eq{Eq:GF}, we find
\be
\E_{lm}(r)= \GF_{lm}^{l'm'}(r,R;k) \bs_{l'p'}.
\label{E-GF}
\ee
Finally, we equate the two forms of the electric field, \Eq{E-SM} and \Eq{E-GF}, at the point $r=R$. Strictly speaking, we equate only their tangent components, in accordance with Maxwell's boundary conditions, which is equivalent to equating their projections onto the polarization vectors ${\bf e}_p$:
\bea
&&\delta_{ll'} \delta_{mm'}{\bf e}_{p}\cdot \E^{\rm in}_{lp'}(R,k)+\sum_{p} S_{lmp}^{l'm'p'}(k) {\bf e}_{p} \cdot\E^{\rm out}_{lp}(R,k)
\nonumber\\
&&=
 \sigma_{l'p'}{\bf e}_{p}\cdot \GF_{lm}^{l'm'}(R,R;k) {\bf e}_{p'}.
\eea
Introducing projections of the GF onto TE and TM polarizations,
\be
{\cal G}_{lmp}^{l'm'p'}(k)={\bf e}_{p}\cdot \GF_{lm}^{l'm'}(R,R;k)  {\bf e}_{p'},
\label{GF-proj}
\ee
and using the explicit form \Eq{norm} of the electric field components at $r=R$,
we arrive at \Eq{SmatrixFinal} which provides the link between the GF and the $S$-matrix.

We note that the established link between the GF and the S-matrix actually allows us to find, via \Eq{E-SM}, the full electromagnetic field outside the basis sphere, from just knowing the GF  on the sphere boundary. This includes both near and far fields. Inside the sphere, the field can be found from \Eq{E-GF} using the GF inside the sphere. Therefore, this approach allows us to determine the electromagnetic field due to a spherical wave excitation of the system at all points of space.

\section{Derivation of the delta source terms replacing incoming spherical waves}
\label{App:Source}
In order to determine the source terms $\bs_{lp}$ in \Eq{ME-source}, rigorously representing the effect of the incoming spherical wave on the
interior region of the sphere $r\leqslant R$, we consider these sources in free space and require that the electric field inside the sphere $r\leqslant R$ is the same as that produced by an incoming wave in vacuum, which is given by Eq.\,(\ref{Eq:TE}) or (\ref{Eq:TM}), with $d={\rm in}$. This electric field is therefore the solution
of the wave equation
\be
-\hat{L}_{l}(r) \E(r) +k^2 \E(r)=\frac{\bs}{R^2}\delta(R-r)\,,
\label{wave-equ-source}
\ee
in which we omitted, for brevity, the indices $l$ and $p$. Below we solve this equation for both TE and TM polarizations, finding the source $\bs$ from the known field $\E(r)$ inside the sphere. As in \App{App:waveequ}, we deal here explicitly with the three components of the vectors of the electric field and the source in VSH space,
which are given by $E_i(r) = [\E(r)]_i$ and $\sigma_i = [\bs]_i$, respectively, with $i=1\,,2\,,3$.

For a TE-polarized wave, having only the component $E_1$, we obtain $\sigma_2=\sigma_3=0$ and from \Eq{wave-equ-source}
\be
\frac{1}{r}\frac{d^2}{dr^2} r E_1 - \frac{\alpha_l^2}{r^2} E_1+k^2 E_1=\frac{\sigma_1}{R^2}\delta(R-r)\,,
\label{we1}
\ee
which, according to \Eq{Eq:TE}, should have the following solution:
\be
E_1(r)=\left\{ \begin{array}{ll}
\tilde{h}_{l,{\rm in}}(r,k) & r\leqslant R\\
\tilde{h}_{l,{\rm out}}(r,k) & r>R\,,
\end{array}
\right.
\ee
continuous at $r=R$ and representing an incoming wave inside and an outgoing wave outside the sphere. Here, the functions $\tilde{h}_{ld}(r,k)$ are defined in \Eq{Eq:tilde_h}.  The discontinuity of the derivative of $E_1$ at $r=R$ determines the source term imitating the incoming wave. This can be found by integrating \Eq{we1}, which leads to
\be
\left.\frac{d E_1}{dr}\right|_{R-0_+}^{R+0_+}=\frac{\sigma_1}{R^2}\,,
\ee
where $0_+$ is a positive infinitesimal. From this we find
\bea
\sigma_1&=&R^2[\tilde{h}'_{l,{\rm out}}(R,k)-\tilde{h}'_{l,{\rm in}}(R,k)]\nonumber\\
&=&R[\gamma^{-1}_{l,{\rm out}}-\gamma^{-1}_{l,{\rm in}}]=\sigma_{l,{\rm TE}}\,,
\label{sigma1}
\eea
using $R\tilde{h}'_{ld}(R,k)=\gamma^{-1}_{ld}-1$ and $\tilde{h}'_{ld}(r,k)\equiv \frac{d}{d r}\tilde{h}_{ld}(r,k)$. Clearly, \Eq{sigma1} defining $\sigma_{l,{\rm TE}}$ coincides with \Eq{Eq:J_TE}.

A TM-polarized wave has two non-vanishing components of the field, $E_2$ and $E_3$. Therefore, in this polarization, $\sigma_1=0$,  and the other two components of the source, $\sigma_2$ and $\sigma_3$, are found from \Eq{wave-equ-source}, which in this case reduces to the following pair of coupled equations:
\bea
\frac{1}{r}\frac{d^2}{dr^2} r E_2 - \frac{\alpha_l}{r} \frac{d}{dr}  E_3+k^2 E_2&=&\frac{\sigma_2}{R^2}\delta(R-r)\,,
\label{we2}
\\
\frac{\alpha_l}{r^2}\frac{d}{dr} r E_2 - \frac{\alpha^2_l}{r^2}E_3+k^2 E_3&=&\frac{\sigma_3}{R^2}\delta(R-r)\,.
\label{we3}
\eea
Its solution should be a TM wave in vacuum, which, according to \Eq{Eq:TM}, is given by
\be
\frac{r}{R}E_2(r)=\left\{ \begin{array}{ll}
\tilde{\xi}_{l,{\rm in}}(r,k) & r\leqslant R \\
\tilde{\xi}_{l,{\rm out}}(r,k) & r>R
\end{array}
\right.
\ee
and
\be
\frac{r}{R \alpha_l}E_3(r)=\left\{ \begin{array}{ll}
\gamma_{l,{\rm in}}\tilde{h}_{l,{\rm in}}(r,k) & r\leqslant R\\
\gamma_{l,{\rm out}}\tilde{h}_{l,{\rm out}}(r,k) & r>R \,,
\end{array}
\right.
\label{TM-E3}
\ee
where the functions $\tilde{\xi}_{ld}(r,k)$ are defined in \Eq{Eq:tilde_h} and the coefficients $\gamma_{ld}$ are given by \Eq{Eq:gamma}.
Again, the obtained solution describes an incoming wave inside the sphere and an outgoing wave outside it. Note that the component $E_2$ is continuous, but $E_3$ is not, as it corresponds to the part of the electric field which is normal to the sphere surface where the delta-like source with a virtual electric currents is placed. Using the continuity of $E_2$ and integrating Eqs.(\ref{we2}) and (\ref{we3}), results in
\be
\frac{1}{R} \left.\left(\frac{d}{dr} r E_2- \alpha_l E_3\right)\right|_{R-0_+}^{R+0_+} =\frac{\sigma_2}{R^2}
\label{sol2}
\ee
and $ \sigma_3=0$\,, respectively. Equation (\ref{sol2}) can be further simplified, using Eqs.\,(\ref{TM-equ2}) and (\ref{TM-E3}) and the fact that $\tilde{h}_{l,{\rm in}}(R,k)=\tilde{h}_{l,{\rm out}}(R,k)=1$. Then we obtain
\be
\sigma_2=\frac{k^2 R^3}{\alpha_l} E_3\bigr|_{R-0_+}^{R+0_+}=k^2 R^3(\gamma_{l,{\rm in}}-\gamma_{l,{\rm out}})=\sigma_{l,{\rm TM}}\,,
\label{sigma2}
\ee
which is the same as \Eq{Eq:J_TM}.

Equations (\ref{sigma1}) and (\ref{sigma2}) thus determine the non-vanishing components of the source terms at $r=R$ which exactly produce the incoming spherical waves in TE and TM polarization, respectively,  normalized according to \Eq{norm}. This result is presented in a compact form by \Eq{sigmas}.

\section{Expansion of a plane wave into vector spherical harmonics}\label{App:PlaneWave}
Consider a linearly polarized plane wave propagating in free space in $z$ direction.  In spherical coordinates, the electric field of a plane wave polarized in $x$ and $y$ direction is given by
\be
\E^x(\r) = \left[ \cos\varphi (\sin\theta\er + \cos\theta\etheta)-\sin\varphi \ephi\right]  e^{ikr\cos\theta}\,,
\ee
\be
\E^y(\r) = \left[ \sin\varphi (\sin\theta\er + \cos\theta\etheta)+\cos\varphi \ephi\right]  e^{ikr\cos\theta}\,,
\ee
respectively, where $\er$, $\etheta$, and $\ephi$ are the unit vectors in spherical coordinates. The corresponding fields $\E^x_{lm}(r)$ and $\E^y_{lm}(r)$ in the VSH basis can be found using the general definition \Eq{coef}, the well-known expansion of a scalar plane wave into scalar spherical harmonics~\cite{LandauLifshitz_3},
\be
e^{ikr\cos\theta}=\sum_{l=0}^\infty i^l \sqrt{4\pi(2l+1)} j_l(kr)Y_{l0}(\Omega)\,,
\ee
and the following explicit form of the VSHs:
\bea
\Y_{1lm}(\Omega)&=& \frac{1}{\alpha_l} \left( -\frac{m}{\sin\theta} Y_{l,-m}
 \etheta + \frac{\partial Y_{lm}}{\partial\theta} \ephi  \right),\\
\Y_{2lm}(\Omega)&=& \frac{1}{\alpha_l} \left(\frac{\partial Y_{lm} }{\partial\theta}
 \etheta +\frac{m}{\sin\theta} Y_{l,-m} \ephi  \right),\\
 \Y_{3lm}(\Omega) &=& Y_{lm}\er\,,
\eea
where $\alpha_l= \sqrt{l(l+1)}$. In particular, we calculate all three components  of the field in each polarization one by one. Consider $[\E^x_{lm}(r)]_1$ for illustration:
\bea
[\E^x_{lm}(r)]_1&=&\int\Y_{1lm}(\Omega)\cdot \E^x(\r)d\Omega
\nonumber\\
&=&\frac{1}{\alpha_l}\sum_{l'=0}^\infty i^{l'} \sqrt{4\pi(2l'+1)} j_{l'}(kr)J_{l'l}^m\,,
\eea
where
\bea
J_{l'l}^m &=&\int Y_{l'0} \left( -m\cos\varphi \frac{\cos\theta}{\sin\theta} Y_{l,-m}
 - \sin\varphi \frac{\partial Y_{lm}}{\partial\theta} \right)  d\Omega
\nonumber\\
&=&\delta_{m,-1}\sqrt{\frac{2l'+1}{2}}\frac{\sqrt{2l+1}}{2\alpha_l} I_{l'l}
\eea
and
\bea
I_{l'l}&=& \!\!\int _0^\pi \!\!\!P_{l'}^0(\cos\theta) \!\left[\frac{\cos\theta}{\sin\theta} P^1_l(\cos\theta) +\frac{\partial P^1_l(\cos\theta)}{\partial\theta} \right]\!\sin\theta d\theta
\nonumber\\
&=&-\frac{2\alpha_l^2}{2l+1} \delta_{ll'}\,.
\eea
In calculating the last integral, we have used the orthogonality of Legendre polynomials $P^m_l(\cos\theta)$ and recursive relations involving their derivatives. Repeating this exercise for the other components, we obtain:
\be
\E^{x,y}_{lm}(r)=-\delta_{m,\mp1} \eta_l \left(\begin{array}{c}
j_l(kr)\\ 0\\0\end{array}\right)\pm
\delta_{m,\pm1} \frac{\eta_l}{kr} \left(\begin{array}{c}
0\\ \zeta'_l(kr)\\\alpha_l j_l(kr)\end{array}\right)
\label{xypol}
\ee
with
\be
 \eta_l=i^l\sqrt{2\pi(2l+1)}
\label{etadef}
\ee
and $\zeta_l(x)=x j_l(x)$. The upper (lower) sign in \Eq{xypol} corresponds to the $x$ ($y$) polarization of the plane wave.

Now, we note that the two vector functions which appear in \Eq{xypol} can be obtained by combining the incoming and outgoing spherical wave in free space, which are given by Eqs.\,(\ref{Eq:TE}) and (\ref{Eq:TM}), respectively, for the TE and TM polarizations. Being represented by the Hankel functions of the first and second kind, these solutions are divergent at the origin but their combination leading to the spherical Bessel functions are regular at $r=0$, as it should be for the plane wave. These unique combinations have the following form:
\bea
\sum_{d={\rm in,out}} \beta^d_{l,{\rm TE}} \E^d_{l,{\rm TE}}(r,k)&=&\left(\begin{array}{c}
j_l(kr)\\ 0\\0\end{array}\right),
\label{TEpol}
\\
\sum_{d={\rm in,out}} \beta^d_{l,{\rm TM}} \E^d_{l,{\rm TM}}(r,k)&=&\frac{1}{kr} \left(\begin{array}{c}
0\\ \zeta'_l(kr)\\\alpha_l j_l(kr)\end{array}\right),
\label{TMpol}
\eea
for TE and TM polarizations, respectively, with
\be
\beta^d_{l,{\rm TE}}= \frac{1}{2} h_l^d(kR)\,,\ \ \ \beta^d_{l,{\rm TM}}= \frac{1}{2}\frac{1}{kR} \xi'_{ld}(kR)\,,
\label{betadef}
\ee
$\xi_{ld}(x)=x h^d_l(x)$, and the fields $\E^d_{lp}(r)$ defined by Eqs.\,(\ref{Eq:TE}) and (\ref{Eq:TM}).
This allows us to apply the expansion \Eq{Eq:EAEY} to a polarized plane wave. Denoting the expansion coefficients $A^{d}_{lmp}$ in the case of a plane wave as $B^{dj}_{lmp}$, where $j=x,y$, we obtain from \Eq{Eq:EAEY}
\begin{equation}
\E^{j}_{lm}(r) = \sum_{pd} B^{dj}_{lmp} \E^d_{lp}(r,k)\,.
\label{Bdef}
\end{equation}
 Comparing \Eq{Bdef} with \Eq{xypol} and using \Eqs{TEpol}{betadef}, we arrive at
\be
B^{dj}_{lmp}=\tau_{mp}^j \eta_l \beta_{lp}^d\,,
\label{Bs}
\ee
where
\bea
&& \tau^x_{m,{\rm TE}}=-\delta_{m,-1}\,,\ \ \ \ \ \tau^x_{m,{\rm TM}}=i\delta_{m,+1}\,,
\\
&& \tau^y_{m,{\rm TE}}=-\delta_{m,+1}\,,\ \ \ \ \ \tau^y_{m,{\rm TM}}=-i\delta_{m,-1}\,,
\eea
and $\eta_l$ and $\beta_{lp}^d$ are given by  Eqs.\,(\ref{etadef}) and (\ref{betadef}), respectively.

\section{Derivation of the equivalent surface current}\label{App:SurfCurr}

Applying the Helmholtz operator to the electric field $\iE_{lm}(r)\Theta(r-R)$ of the first part of the system introduced in \Sec{Sec:GF_IP}, using the explicit expression for the double curl operator \Eq{Eq:L}, and noting that $\iE_{lm}(r)$ is a solution of \Eq{ME-mapped} in the region $|\r|>R$, we obtain
\bea
&&\sum_{l'm'} \LL_{lm}^{l'm'}(r;k) \iE_{l'm'}(r)\Theta(r-R)
\nonumber \\
&=&(k^2 - \hat{L}_l)\iE_{lm}(r)\Theta(r-R)
\nonumber \\
&=&\Theta(r-R)(k^2 - \hat{L}_l)\iE_{lm}(r)
\nonumber \\
&&+\J_{lm}\frac{\delta(r-R)}{r^2}+\Q_{ilm}\delta'(r-R),\label{Eq:EG}
\eea
where
\be
\frac{\J_{lm}}{R}\!=\!
\begin{pmatrix}
R[\iE'_{lm}(R)]_1+2[\iE_{lm}(R)]_1\\
R[\iE'_{lm}(R)]_2+2[\iE_{lm}(R)]_2-
\sqrt{l(l+1)}[\iE_{lm}(R)]_3\\
\sqrt{l(l+1)}[\iE_{lm}(R)]_2
\end{pmatrix}
\ee
and
\begin{gather}
\Q_{lm}=
\begin{pmatrix}
[\iE_{lm}(R)]_1\\
[\iE_{lm}(R)]_2\\
0
\end{pmatrix}.
\end{gather}
$\Q_{lm}$ vanishes under the condition \Eq{Eq:Econd}, and using \Eq{Eq:EG} and \Eq{Eq:EG_iE} we find that the second part of the electric field $\E^\mathrm{G}_{lm}(r)$ satisfies equation \Eq{Eq:EG_2}.

Note that the first and the second components of the electric field $\E_{lm}(r)$ are continuous functions of the coordinate $r$ at the surface $r=R$.
Indeed, since the first and the second components of the field $\iE_{lm}(r)$ are continuous and vanish at $r=R$ due to \Eq{Eq:Econd}, multiplying them with the Heaviside step function $\Theta(r-R)$ retains their continuity.
The first two components of the field $\E^\mathrm{G}_{lm}(r)$ are continuous as well, since otherwise their second derivatives (see the explicit form \Eq{Eq:L} of the operator $\hat{L}_l$) would result in the derivative of the Dirac delta function $\delta'(r)$, which is not present in the right hand side of \Eq{Eq:EG_2}.

\section{Scattering and absorption  cross-sections}\label{App:SC}
The scattering cross-section is defined as the area orthogonal to the propagation direction of the  plane-wave excitation transmitting the same power as is scattered by the system under this excitation. Therefore, we start by considering an open system illuminated by a plane wave. The incoming amplitudes $A^\mathrm{in}_{plm}$ [see \Eqs{Eq:EAEY}{SmatrixDef}] are then given by the incoming plane wave expansion coefficients \Eq{etadef} for linear  or circular polarization:
\begin{equation}
A^\mathrm{in}_{lmp} = B^{\mathrm{in}}_{lmp}.\label{Eq:SCS_A_in}
\end{equation}
In particular, \Eq{Bs} in \App{App:PlaneWave} gives the explicit form of the expansion coefficients $B^{dj}_{lmp}$ for linear polarized plane waves (here $j=x$ or $y$).
The outgoing amplitudes $A^\mathrm{out}_{lmp}$ instead are given by the outgoing plane wave expansion coefficients $B^{\mathrm{out}}_{lmp}$ plus the plane wave scattering amplitudes $A^\mathrm{sca}_{lmp}$ of the system
\begin{equation}
A^\mathrm{out}_{lmp} = B^{\mathrm{out}}_{lmp} + A^\mathrm{sca}_{lmp}.\label{Eq:SCS_A_out}
\end{equation}
Using the scattering matrix, which is connecting incoming and outgoing amplitudes, by substituting \Eqs{Eq:SCS_A_in}{Eq:SCS_A_out} into \Eq{SmatrixDef}, we find the plane wave scattering amplitudes as
\begin{equation}
A^\mathrm{sca}_{plm} = \sum_{l'm'p'} S^{l'm'p'}_{lmp} B^{\mathrm{in}}_{l'm'p'} - B^{\mathrm{out}}_{lmp}.
\end{equation}

To determine the scattering cross-section $\sigma^\mathrm{sca}$, we then calculate the electromagnetic power scattered by the system in the far field, normalizing it to the power flux $S_0={c}/{(8\pi)}$ of the incoming plane wave (note that we use here a unity plane wave field amplitude for brevity, as the result is independent of this amplitude):
\begin{equation}
\sigma^\mathrm{sca}=\frac{1}{S_0}\lim\limits_{r\rightarrow\infty} r^2 \int  \er\cdot\S^\mathrm{sca}(r,\Omega)d\Omega.\label{Eq:SigmaSCA}
\end{equation}
Here, $\S^\mathrm{sca}(r,\Omega)$ is the Pointing vector of the scattered field
\begin{equation}
\S^\mathrm{sca} = \frac{c}{8\pi} \Re\left[ \E^\mathrm{sca} \times (\H^\mathrm{sca})^\ast \right]. \label{Eq:PointVec}
\end{equation}

Using Maxwell's equations, one can show that the magnetic field of the spherical electromagnetic waves \Eqs{Eq:TE}{Eq:TM} is given by
\bea
\H^d_{l,\mathrm{TE}}(r)&=&
i k R \gamma_{ld} \E^d_{l,\mathrm{TM}}(r),\label{Eq:H_TM}\\
\H^d_{l,\mathrm{TM}}(r)&=&
i (k R \gamma_{ld})^{-1} \E^d_{l,\mathrm{TE}}(r).\label{Eq:H_TE}
\eea

Both the electric and magnetic fields of the outgoing spherical electromagnetic waves contain spherical Hankel functions of the first kind, which for a real argument $x$ has the asymptotic behaviour
\begin{equation}
h^{(1)}_l(x) \approx (-i)^{l+1}\frac{e^{ix}}{x} \quad \mathrm{for} \quad x\gg l.
\end{equation}
Using this asymptotics in \Eqs{Eq:TE}{Eq:TM} and \Eqs{Eq:H_TM}{Eq:H_TE}, substituting the electric and magnetic fields into the Pointing vector \Eq{Eq:PointVec}, and the result into \Eq{Eq:SigmaSCA}, we obtain the total scattering cross-section
\begin{equation}
\sigma^\mathrm{sca}=\sum_{lmp}\Gamma_{lmp} \left| A^\mathrm{sca}_{lmp} \right|^2,
\label{Eq:sscat}
\end{equation}
where
\bea
\Gamma_{lm,{\rm TE}} &=& \left| kh^{(1)}_l(k R) \right|^{-2}\,,\\
\Gamma_{lm,{\rm TM}} &=& k^2 R^2 | \gamma_{l,{\rm out}} |^2 \Gamma_{lm,{\rm TE}}\,.
\eea
In deriving this expression, we have used the fact that all $\Y_{3lm}(\Omega)$ are parallel to $\er$ [see \Eq{Eq:Y3}], so that their cross product with any VSH has vanishing projection along $\er$.
Furthermore, using the relation $\left[\mathbf{a}\times(\mathbf{b}\times\mathbf{c})\right]=\mathbf{b}(\mathbf{a}\cdot\mathbf{c})-\mathbf{c}(\mathbf{a}\cdot\mathbf{b})$, and the definition of the VSHs \Eqs{Eq:Y1}{Eq:Y3}, we find
\begin{equation}
\int \Y_{2lm}(\Omega) \times \Y_{1l'm'}(\Omega)d\Omega=\er\delta_{ll'}\delta_{mm'}\,,
\end{equation}
leading to the simple diagonal expression in \Eq{Eq:sscat}.

The absorption cross-section $\sigma^\mathrm{abs}$, which is defined equivalent to the scattering cross-section, but for the power absorbed by the system, can be calculated as the difference between the power flowing inwards and outwards, normalized to the power flux density of the plane wave, yielding
\begin{equation}
\sigma^\mathrm{abs} = \sum_{lmp}\Gamma_{lmp} \left( \left| B^\mathrm{in}_{lmp} \right|^2 -
\left| A^\mathrm{out}_{lmp} \right|^2 \right)\,,\label{Eq:sabs}
\end{equation}
which is derived similar to \Eq{Eq:sscat}.

\section{First-order treatment of basis extension}\label{App:TruncatedSpace}

Here we show how an extended basis of the RSE can be taken into account in first order with a reduced numerical complexity. Suppose that in order to find the RS wave numbers $\varkappa_\nu$ of the new system, as well as the expansion coefficients $C_{n\nu}$ of the RS fields, see \Eq{Exp}, we truncate the infinite matrix eigenvalue problem \Eq{Eq:EigenvalueProblem} of the RSE, keeping a finite number $N+L$ of RSs in the basis. Here $N$ is the number of RSs  taken into account exactly and having the wave numbers $|k_n|<\kmax$, while $L$ is the number of additional RSs with higher wave numbers $\kmax\le|k_n|<\kmax'$, giving the basis extension, which will be taken into account in first order. To dicuss the method, we write \Eq{Eq:EigenvalueProblem} in matrix form with $(N+L)$ dimensional square matrices $\mk$ and $\mM$ and an $(N+L)$ dimensional eigenvector $\bc_\nu$
\begin{equation}
\mathbbm{k} \bc_\nu = \varkappa_\nu \mathbbm{M} \bc_\nu\,.
\label{mequ}
\end{equation}
 Here, $\mk$ is a diagonal matrix containing the wave numbers of the basis RSs sorted in ascending order. We now split the notation explicitly into the $N$ and $L$ RSs, so that the matrices $\mk$ and $\mM$ split into four sub-matrices, with $N$-dimensional square top-left sub-matrices $\mk_0$ and $\mM_{00}$, and the vector $\bc_\nu$ splits into two sub-vectors, with an $N$-dimensional top sub-vector $\bc_{0\nu}$. The matrix equation (\ref{mequ}) accordingly reads
\begin{equation}
\begin{pmatrix}
\mk_0 & 0\\
0 & \mk_1
\end{pmatrix}
\begin{pmatrix}
\bc_{0\nu}\\
\bc_{1\nu}
\end{pmatrix}
= \varkappa_\nu
\begin{pmatrix}
\mM_{00} & \mM_{01}\\
\mM_{10} & \mM_{11}
\end{pmatrix}
\begin{pmatrix}
\bc_{0\nu}\\
\bc_{1\nu}
\end{pmatrix},\label{Eq:LargeEig}
\end{equation}
where $\mM_{00}$ and $\mM_{11}$ are symmetric matrices, and  $\mM_{01}$ is the transpose of $\mM_{10}$. Neglecting the $L$ RSs,  \Eq{Eq:LargeEig} reduces to the $N \times N$ matrix equation
\begin{equation}
\mk_0
\bc_{0\nu}
= \varkappa_\nu
\mM_{00}
\bc_{0\nu}\,.\label{Eq:SmallEig}
\end{equation}
Its solution is the RSE result presented in \Sec{Sec:Results}.
Since the time required for solving an eigenvalue problem scales with the third power of its size, the computational complexity of \Eq{Eq:SmallEig} is $(1+L/N)^3$ times lower than  \Eq{Eq:LargeEig}.

We now find the lowest-order approximation for the eigenvector component $\bc_{1\nu}$ for the given state $\nu$. This correction to the RS fields is proportional to the perturbation matrix $\mM_{10}$. Indeed, neglecting all off-diagonal elements of $\mM_{11}$, we obtain from \Eq{Eq:LargeEig}
\begin{equation}
\left(\mk_1-\varkappa_\nu\mD_{11}\right) \bc_{1\nu} = \varkappa_\nu \mM_{10} \bc_{0\nu}\,,\label{Eq:MatrE1}
\end{equation}
where  $\mD_{11}$ is the diagonal part of $\mM_{11}$, and $\bc_{0\nu}$ and $\varkappa_\nu$ are the solutions of \Eq{Eq:SmallEig}.
The computational complexity of \Eq{Eq:MatrE1} is of order $LN^2$, with a prefactor which we found to be about 10 times lower than the matrix diagonalization used for solving \Eq{Eq:SmallEig}. The first-order treatment thus allows us, for the same compute time, to extend the basis about 10-fold, which, if used directly in \Eq{Eq:SmallEig}, would require a 1000-fold increased compute time.

In the refined RSE, we include in the expansion of the RS fields $\e_\nu$ given by \Eq{Exp} the components of the second part of the basis RSs, with the expansion coefficients $\bc_{1\nu}$ given by \Eq{Eq:MatrE1}.

We now evaluate the lowest-order correction $-\delta\varkappa_\nu$ to the eigenvalues $\varkappa_\nu$ of the first subgroup, given by \Eq{Eq:SmallEig}, due to the basis extension and the coefficients $\bc_{1\nu}$ determined in first order via \Eq{Eq:MatrE1}. We find from~\Eq{Eq:LargeEig}
\begin{equation}
\mk_0
\bc_{0\nu} = (\varkappa_\nu-\delta\varkappa_\nu)(\mM_{00} \bc_{0\nu}+ \mM_{01} \bc_{1\nu})\,,
\end{equation}
and using \Eq{Eq:SmallEig} obtain
\be
\delta \varkappa_\nu\mM_{00} \bc_{0\nu}= \varkappa \mM_{01} \bc_{1\nu}\,,
\ee
where we have neglected the higher-order term $\delta\varkappa_\nu\mM_{01} \bc_{1\nu}$. Finally, taking the dot product of the above equation with $\bc_{0\nu}$, and using the normalization $\bc_{0\nu}\cdot\mM_{00}\bc_{0\nu}=1$ given by \Eq{Eq:Normalization_3} up to first order in $\mM_{01}$, we arrive at the second-order correction to the wave-numbers:
\begin{equation}
\delta\varkappa_\nu = \varkappa_\nu (\bc_{1\nu}\cdot \mM_{10} \bc_{0\nu})\,,
\label{Eq:deltaKappa}
\end{equation}
where $\bc_{0\nu}$ and $\bc_{1\nu}$ are approximated by \Eq{Eq:SmallEig} and  \Eq{Eq:MatrE1}, respectively.
In numerical calculations, we used this equation only for estimation of the eigenvalues error due to the following reason.
After solving \Eq{Eq:SmallEig}, one can find that some wave-number eigenvalues can have the absolute value of the imaginary part smaller than the error given by \Eq{Eq:deltaKappa} or can even have a positive imaginary part, which is not physical.
In our calculation, the wave numbers $\varkappa_\nu$ for such modes are replaced by $\Re\varkappa_\nu - i|\delta\varkappa_\nu|$, where $|\delta\varkappa|$ is determined by \Eq{Eq:deltaKappa}.
\bigskip
\section{Analytical solution for the scattering matrix of a dielectric sphere}
\label{App:AnalyticalSolution_S}

The elements of the scattering matrix are found by using Maxwell's boundary conditions for the analytic solutions \Eqs{Eq:TE}{Eq:TM} on the sphere boundary, leading to
\begin{gather}
S_{lm,\mathrm{TE}}^{\rm (exact)}(k) =
-\frac{\tilde{j}_{l}'(R,k)-\tilde{h}'_{l,{\rm in}}(R,k)}{\tilde{j}'_{l}(R,k)-\tilde{h}'_{l,{\rm out}}(R,k)},\\
S_{lm,\mathrm{TM}}^{\rm (exact)}(k) =
-\frac{\varepsilon \tilde{\gamma}_{l}(k)-\gamma_{l,{\rm in}}(k)}{\varepsilon \tilde{\gamma}_{l}(k)-\gamma_{l,{\rm out}}(k)},
\end{gather}
where the prime means the first derivative with respect to the first argument. The functions $\tilde{h}'_{ld}(r,k)$ and $\gamma_{ld}(k)$ are defined by \Eqs{Eq:tilde_h}{Eq:gamma}, and
\be
\tilde{j}_l(r,k)= \frac{j_l(nkr)}{j_l(nkR)}\,, \quad
\tilde{\gamma}_{l}(k) = \frac{j_l(nkR)}{\zeta'_l(nkR)}.
\ee
Here, $j_l(x)$ is the spherical Bessel function and $\zeta_l(x) = x j_l(x)$.

The above equations for the S-matrix yield the scattering cross-sections of the Mie theory, which can be found e.g. in \citep{BohrenBook98}.

\end{document}